\documentclass[fleqn,usenatbib]{mnras}

\usepackage{newtxtext,newtxmath}

\usepackage[T1]{fontenc}

\DeclareRobustCommand{\VAN}[3]{#2}
\let\VANthebibliography\thebibliography
\def\thebibliography{\DeclareRobustCommand{\VAN}[3]{##3}\VANthebibliography}

\usepackage{graphicx}	
\usepackage{amsmath}	
\usepackage{longtable}  
\usepackage{multirow}
\usepackage{manyfoot}
\usepackage{lipsum}
\usepackage{subcaption}
\usepackage{deluxetable}
\usepackage{placeins}
\usepackage{xcolor}

\defcitealias{kini:2023}{K23}

\DeclareNewFootnote{A}[arabic]
\DeclareNewFootnote{B}[alph]

\title[Pulse Profile Modelling of Thermonuclear Burst Oscillations II]{Pulse Profile Modelling of Thermonuclear Burst Oscillations II: Handling variability}

\author[Kini~et~al.]{Yves~Kini$^{1}$
\thanks{E-mail: \href{mailto:y.kini@uva.nl}{y.kini@uva.nl}},
Tuomo~Salmi$^{1}$, 
Serena~Vinciguerra$^{1}$,
Anna~L.~Watts$^{1}$,
Devarshi~Choudhury$^{1}$,\newauthor  
Slavko~Bogdanov$^{2}$, 
Johannes Buchner$^{3}$,
Zach~Meisel$^{4}$ and
Valery~Suleimanov$^{5}$ 
\\
$^{1}$Anton Pannekoek Institute for Astronomy, University of Amsterdam, Science Park 904, 1090GE Amsterdam, the Netherlands\\
$^{2}$Columbia Astrophysics Laboratory, Columbia University, 550 West 120th Street, New York, NY 10027, USA\\
$^{3}$Max Planck Institute for Extraterrestrial Physics, Giessenbachstrasse, 85741 Garching, Germany\\
$^{4}$Institute of Nuclear \& Particle Physics, Department of Physics \& Astronomy, Ohio University, Athens, Ohio 45701, USA\\
$^{5}$Institut f\"ur Astronomie und Astrophysik, Kepler Center for Astro and Particle Physics, Universit\"at T\"ubingen, Sand 1, D-72076 T\"ubingen, Germany
}

\date{Accepted XXX. Received YYY; in original form ZZZ}

\pubyear{20XX}

\begin{document}
\label{firstpage}
\pagerange{\pageref{firstpage}--\pageref{lastpage}}
\maketitle


\begin{abstract}

Pulse profile modelling is a relativistic ray-tracing technique that can be used to infer masses, radii and geometric parameters of neutron stars. In a previous study, we looked at the performance of this technique when applied to thermonuclear burst oscillations from accreting neutron stars.  That study showed that ignoring the variability associated with burst oscillation sources resulted in significant biases in the inferred mass and radius, particularly for the high count rates that are nominally required to obtain meaningful constraints.  In this follow-on study, we show that the bias can be mitigated by slicing the bursts into shorter segments where variability can be neglected, and jointly fitting the segments. Using this approach, the systematic uncertainties on the mass and radius are brought within the range of the statistical uncertainty. With about 10$^6$ source counts, this yields uncertainties of approximately 10\% for both the mass and radius. However, this modelling strategy requires substantial computational resources. We also confirm that the posterior distributions of the mass and radius obtained from multiple bursts of the same source can be merged to produce outcomes comparable to that of a single burst with an equivalent total number of counts.\

\end{abstract}



\section{Introduction}\label{sec:intro}

Astrophysics and nuclear physics have long sought to understand the fundamental interactions of matter at low temperatures and high densities. Neutron stars (NSs) exclusively hold the key to this goal since their cores are expected to be extremely dense and cold (compared to the Fermi temperature). A promising approach to studying the properties of the core is to use measurements of their masses and radii to infer the equation of state (EoS) that governs the matter inside their interiors \citep[see e.g.][]{ Lattimer:2012, Oertel:2016, Baym:2017, Tolos:2020, Yang:2019, Hebeler:2020}.

The masses, radii, geometric parameters and surface hot spot patterns of neutron stars can be deduced by using the technique of Pulse Profile Modelling (PPM) to analyze the X-ray pulsations originating from the surface of X-ray pulsars \citep{Pechenick:1983,Chen:1989,Page:1995,Miller:1998,Braje:2000, Weinberg:2001,Beloborodov:2002, Poutanen:2006, Cadeau:2007,Morsink:2007, Baubock:2012,Lo:2013ava, Psaltis:2013fha, Miller:2015,Stevens:2016, Nattila:2018, Bogdanov:2019}. This technique, based on relativistic ray-tracing, takes advantage of the fact that the observed X-ray photons encode data about the source's properties and the interactions that the photons experienced en route to the observer.  To decrypt the encoded data and infer the relevant stellar properties, PPM uses both mathematical models (e.g. of the space-time, the hot spot properties, and the atmosphere) and statistical inference tools.

Knowledge of several simultaneous masses and radii is required to establish the EoS precisely \cite[see e.g.][]{Ozel:2009b,Ozel:2016} and by extension, the state of matter prevailing in the core. Applying PPM\footnote{Alternative methods, such as phase-averaged spectral modelling of bursters and quiescent Low Mass X-ray Binaries (LMXBs) \citep[see e.g.][]{Ozel:2009, Guver:2010b,Suleimanov:2011, Steiner:2013, Guillot:2014, Nattila:2016, Steiner:2018} can also be used to derive the mass and radius. See \citet{Degenaar:2018} for a review of this technique including a discussion of model dependencies and potential uncertainties.} to data from the Neutron Star Interior Composition Explorer \citep[NICER;][]{NICER}, masses and radii have been so far inferred for two Rotation-Powered Millisecond Pulsars (RMPs) - PSR J0030$+$0451 \citep[][]{Riley:2019, Miller:2019} and PSR J0740$+$6620 \citep{Riley:2021, Miller:2021, Salmi:2022} - with results for more RMPs to come \citep[see][for a non-exhaustive list of targeted sources]{Bogdanov:2019a}. For RMPs, the pulsations are stable, which means that temporal variability can be neglected.  However there is uncertainty over the size, shape and temperature distribution of the hot spots, which originate as electrons accelerated in the magnetic field bombard the stellar surface, heating the magnetic poles. This is addressed in the modelling using simplified parameterized models that try to capture a wide range of possibilities, motivated by current pulsar theory. 

PPM can in principle be extended to other categories of pulsating neutron stars \citep[][]{Watts:2016}, in particular Accretion-powered Millisecond Pulsars \citep[AMPs; see e.g.][]{Salmi:2018} and Thermonuclear Burst Oscillation (TBO) sources \citep[see e.g][]{Bhattacharyya:2004pp}. By extending the scope of PPM to encompass other neutron star classes, one obtains a larger sample of inferred masses and radii (reducing statistical uncertainties in inferring EoS parameters), and using a different population allows for independent cross-checks. In this paper we focus on TBO sources.

A thermonuclear burst is a sudden and intense release of X-rays that happens in a neutron star's outer layers. It is caused by a runaway nuclear fusion process in which hydrogen and helium in material accreted from a companion star burn unstably and rapidly, forming heavier elements \citep[for a recent review, see][]{Galloway:2020}. During some bursts, uneven heat distribution on the NS surface or in their atmosphere leads to pulsations in the light curve. These pulsations are commonly referred to as TBOs. The cause of the uneven surface temperature distribution that generates the TBOs is still unclear. Various models, mostly complementary, with their respective strengths and weaknesses, have been proposed to explain this phenomenon. These include: dynamics of the flame front \citep{Strohmayer:1997, Spitkovsky:2002,Cavecchi:2013,Cavecchi:2015, Cavecchi:2016},
buoyant r-modes \citep{Heyl:2004, Lee:2004, Piro:2005,Chambers:2019, Chambers:2020}, a cooling wake \citep{Mahmoodifar:2016} and convective patterns \citep{Garcia:2018}. On top of the uncertainties surrounding the origin of the oscillations, the surface conditions during a burst are chaotic and prone to short-term variability (which is both theorized and observed). There may be variations in the temperature of the star and/or hot spot, and the spot may move or change location. 

This means that unlike for the RMPs, we have to contend with both increased surface temperature distribution uncertainty (due to the wider range of models) and variability.  However TBO sources spin more rapidly than most RMPs, which should help to disentangle the correlation between mass and radius \citep{Lo:2013ava,Psaltis:2013fha}. Moreover, given that bursts are bright events, obtaining the number of counts required to place meaningful constraints on mass and radius is feasible, particularly if data from different bursts can be combined. Previous studies have shown that we would need $\gtrsim 10^6$ counts (from a hot spot) to derive useful constraints on the mass and radius \citep[][]{Lo:2013ava, Psaltis:2013fha}.  This should be easily achievable with proposed large-area X-ray spectral-timing telescopes such as the enhanced X-ray Timing and Polarimetry mission \citep[e-XTP;][]{Zhang:2019, Watts:2019_extp}, and the Spectroscopic Time-Resolving Observatory for Broadband Energy X-rays \citep[STROBE-X;][]{Ray:2019}.

To better understand the impact of the uncertainties regarding the origin of TBOs and the complexities linked to variability, \citet{kini:2023} (hereafter \citetalias{kini:2023}) developed phenomenological models to mimic the observed TBOs in the bursts from the accreting pulsar XTE~J1814$-$338 (hereafter J1814) \citep[][]{Strohmayer:2003}. This particular source has always been regarded as one of the most promising TBO sources for PPM \citep[][]{Bhattacharyya:2004pp} due to its stable pulsations, combined with high root mean square fractional amplitude (rms FA)\footnote{In the literature, the rms FA is denoted in different ways. In e.g. \citet{Edelson} it is called fractional variability (Fvar) while in e.g. \citet{Vaughan} it is referred to as fractional rms variability amplitude.} and harmonic content \citep[see e.g.][]{Braje:2000, Poutanen:2006, Lo:2013ava,Psaltis:2013fha}. Using these phenomenological models, \citetalias{kini:2023} produced synthetic bursts that incorporate time-dependent parameters and conducted parameter recovery disregarding these temporal variabilities. It was found that not appropriately accounting for the variability leads to a bias (i.e. the inferred masses and radii deviate from the injected values due to primarily systematic rather than statistical uncertainty) in estimating the mass and radius. This bias becomes noticeable when the count number reaches 10$^6$ (the threshold for obtaining meaningful constraints, see above).

In this second paper, we investigate methods that can correct the biases introduced by neglecting temporal variability for high counts and accurately determine the parameters of the burster. We also aim to determine the computational cost associated with such methods, to estimate the appropriate allocation of computing resources when analyzing current and future data. To accomplish this, we use a subset of the synthetic data generated in \citetalias{kini:2023}, which encompasses diverse phenomenological models incorporating different variabilities. Finally, we explore strategies to combine knowledge acquired from several individual bursts to achieve more refined constraints for a given source. To achieve this, we employ synthetic data generated using a single phenomenological model that incorporates diverse forms of temporal variability for the time-dependent parameters.  
Since one goal of \citetalias{kini:2023} and the present paper is to provide a foundation to determine the properties of J1814 using the available Rossi X-Ray Timing Explorer \citep[RXTE;][]{Jahoda:1996} Proportional Counter Array (PCA) data, we used RXTE's response matrix throughout these papers.

The rest of this paper is organized as follows. In Section \ref{sec:method}, we provide a brief summary of the necessary model ingredients in our PPM framework. This is followed by a description of the simulated data used and an explanation of our inference procedure, including how we combine information obtained from individual bursts. The next section (Section \ref{sec:result}) presents our findings, which are then discussed along with their implications in Section \ref{sec:discussion}. The closing remarks are presented in Section \ref{sec:Conclusion}. 
\section{METHOD }\label{sec:method} 
As stated in Section \ref{sec:intro}, determining the properties of a NS  using its X-ray pulses requires both mathematical models and statistical tools. The first mathematical model is that for the star's shape and the space-time in which it is embedded. The surface pattern and atmosphere models, which constitute the second component of the mathematical model, determine the source of X-ray photons by taking into account various factors, including the shape of the hot spot(s), the energy and angular distribution (beaming function) of the X-ray photons, and the ways in which the photons interact after they are emitted. 
The third component is an Interstellar Medium (ISM) model, to account for the absorption of photons by the ISM prior to detection. We also need to decide how to treat the presence of background photons from instrument or other sources in the field of view of the NS. Finally, we need to have a model of the instrument's response (RSP), and the uncertainty on that response. Each of these model components must then be incorporated into a simulation pipeline to make computation possible.

While all of the model components mentioned above are described in greater detail in \citetalias{kini:2023}, we provide an overview of the most critical aspects in the first part of this section. The second part, starting from subsection \ref{sec:combinig_formalism} details the formalism for integrating information from separate bursts to derive a more robust estimation of stellar properties.

\subsection{PPM ingredients}

We used the X-ray Pulse Simulation and Inference \citep[\texttt{X-PSI}\footnote{\url{https://github.com/xpsi-group/xpsi.git}};][]{Riley2023} code, version v0.7.9\footnote{With small modifications explained in \citetalias{kini:2023} and in \citet{zenodo_paper1}} to generate data that replicates the properties of the J1814 light curves, which we then use for parameter inference. In \texttt{X-PSI}, it is presumed that the star has an oblate shape and is embedded in a Schwarzschild space-time. When calculating the observed flux, Doppler terms are included to account for rotational effects. Photons emitted on the oblate surface are conveyed to the observer by means of relativistic ray tracing \citep[for details of the oblate Schwarzschild plus Doppler approximation see][]{Cadeau:2007, Morsink:2007,AlGendy:2014, Bogdanov:2019}.

We assumed, as the origin of the pulsations, a single circular X-ray hot spot emitting uniformly. Its main characteristics are: co-latitude ($\Theta_\mathrm{spot}$), phase ($\phi_\mathrm{spot}$), angular radius ($\zeta_\mathrm{spot}$), and temperature ($T_\mathrm{spot}$). We assumed the co-latitude and phase to be constant throughout a burst, but allowed the angular radius and/or temperature of the hot spot to vary depending on the phenomenological model (see Table \ref{tab:models} for details of the varying components of each specific phenomenological model). In some cases, the remaining portion of the star ($T_\mathrm{star}$) was also permitted to emit. We selected models with parameter vectors capable of reproducing the properties of the bursts and TBOs of J1814, as observed with the RXTE PCA: $\sim 10^5$ counts per burst, and $\sim$ 10\% rms FA throughout the burst \citep{Strohmayer:2003}. Table \ref{tab:models} provides a summary of all phenomenological models and other model components necessary for generating synthetic data and performing parameter inference.

\begin{table*}
\begin{minipage}{1.\textwidth}
\renewcommand*{\thempfootnote}{\fnsymbol{mpfootnote}}
\begin{tabular}{lllll}

\hline \hline
 Models &                                                 & $T_{\mathrm{spot}}$                & $\zeta_{\mathrm{spot }}$             & $T_{\mathrm{star}}$ \\ \hline  
\multirow{9}{3cm}{Surface patterns or phenomenological model }    
                       &ST-$\tilde{H}_{T}$                & varying during the burst\footnote{The temperature evolution of GS 1826-24 was used as a basis for the temperature evolution for these models. The average temperature evolution of GS 1826-24 denoted $T_\mathrm{ma6}(t)$, was obtained by averaging the temperatures of the simulated bursts in the ma6 model sequence \citep[see][]{Meisel:2018rsy}, with the exception of the first and last bursts. \label{mesa}}  & not varying during the burst         & --      \\ 
                       &ST-$\tilde{H}_{T,R}$              & varying during the burst\footref{mesa} & varying during the burst\footnote{The angular radius was fine-tuned to replicate the properties of J1814. See Figure 2 in \citetalias{kini:2023} for its evolution. \label{R_t}}             & --     \\ 
                       &ST-$\bar{S}\tilde{H}_{T}$         & varying during the burst\footref{mesa} & not varying during the burst          & not varying during the burst  \\
                       &ST-$\bar{S}\tilde{H}_{T,R}$       & varying during the burst\footref{mesa} & varying during the burst\footref{R_t} & not varying during the burst       \\
                       &ST-$\tilde{S}_{T}\tilde{H}_{T}$   & varying during the burst\footnote{The temperature was fine-tuned to replicate the properties of J1814. See Figure 2 in \citetalias{kini:2023} for its evolution. \label{T_t}}                     & not varying during the burst          &  varying during the burst\footref{mesa} \\ 
                       &ST-$\tilde{S}_{T}\tilde{H}_{T,R}$ & varying during the burst\footref{T_t}  & varying during the burst\footref{R_t} & varying during the burst\footref{mesa}\\
                       &ST-$\tilde{S}_{T}\bar{H}$         & not varying during the burst           & not varying during the burst          & varying during the burst\footref{mesa} \\
                       &ST-$\tilde{S}_{T}\tilde{H}_{R}$   & not varying during the burst           & varying during the burst\footref{R_t} & varying during the burst\footref{mesa} \\ \hline 
\multicolumn{1}{|l|}{Atmosphere} & \multicolumn{4}{l|}{Solar abundance atmosphere for bursting neutron star atmosphere developed by \citet{Valery2012}} \\ \hline 

\multicolumn{1}{|l|}{Background} &  \multicolumn{4}{l|}{Blackbody + power-law: $T_\mathrm{bb}$=0.95 keV, $R_\mathrm{bb}$=1.6 km, $\gamma_\mathrm{pl}$=1.41, A$_\mathrm{pl}$= $3.32\times10^{-2}$ photons $\mathrm{keV}^{-1} \mathrm{cm}^{-2} \mathrm{s}^{-1}$\citep[see][]{Krauss:2005sj} }  \\ \hline 

\multicolumn{1}{|l|}{ISM} & \multicolumn{4}{l|}{ \texttt{TBabs}: this model only considers neutral gas absorption and adopts photoelectric absorption cross-sections from \citet{Wilms:2000}.} \\ \hline 

\multicolumn{1}{|l|}{RSP } & \multicolumn{4}{l|}{RXTE's response matrix (ObsID: 80418-01-02-00 which corresponds to the observation of J1814's burst 10)} \\ \hline

\end{tabular}%

\caption{Summary of the model properties used to produce data and - for all components apart from the background - to conduct inference. No assumption is made about the functional form of the background during the inference. We applied a flat prior distribution for each background variable centered around the true value (for each energy channel), with a hard cut at $\pm 3\sigma$ to minimise any possible biases coming from the inferred background. See Section 2.4 of \citetalias{kini:2023} for more details of the background modelling and  Appendix B.2 \citep{riley_thesis} for the background marginalization procedure. To generate the data, we kept the background count rate at 80 counts/s during the entire burst.}
\label{tab:models}
\end{minipage}
\end{table*}

\subsection{Synthetic data}

In this analysis\footnote{In both \citetalias{kini:2023} and this paper, the assumed spin frequency of the star is 314.0 Hz, the known burst oscillation spin frequency of J1814.}, we used two data subsets generated using the ingredients described in the previous subsection.

Our first data set, which we used to test methods to mitigate bias, is the $10^6$ count \textit{variability data-set} generated in \citetalias{kini:2023}. The parameter vectors in this data set encompass a wide variety of injected values of e.g. mass, radius. For each phenomenological model and parameter vector (3 per model, see  Table A1 of \citetalias{kini:2023}), 10 bursts of $10^5$ counts were generated, each with different random noise injections, as described in that paper.
The 10 bursts were then summed to create a single mega-burst with a total of $10^6$ counts (see an example of a burst in Figure \ref{fig:burst}). 

In this approach, each burst in a given set has exactly the same hot spot properties.  However this may not be realistic.  For J1814,  
the temperature evolution does vary from burst to burst, as indicated by the light curves \footnote{See \citet{Galloway:2020} or \url{https://burst.sci.monash.edu/}.}, and the ignition location and the position of the oscillation hot spot may be subject to variation during and between bursts \citep{Watts:2008,Cavecchi:2022}. 

To explore this more thoroughly, we also generated a second data subset. We used the first parameter vector (s1) of the most complex phenomenological model, ST-$\tilde{S}_{T}\tilde{H}_{T,R}$ (see table A1 in the appendix of \citetalias{kini:2023}).  We generated 9 bursts (whose total counts summed to $\sim 10^6$). 
The mass, radius, distance, inclination, and hydrogen column density were kept constant, but each burst had its own unique evolution (temporal dependence) of  hot spot temperature, hot spot angular radius, and star temperature 
 (see Figure \ref{fig:profiles} of Appendix \ref{sec:appendix}).  In addition, the hot spot's co-latitude and phase varied from burst to burst but remained constant during each one. Each of these nine bursts shared the main features of the J1814 bursts, mainly: $\sim 10^5$ counts and $\sim$ 10\% rms FA throughout the burst. This set of nine bursts is the second data subset\footnote{The bursts differ much more from each other than the bursts generated using this model in the first data subset.}.

\subsection{Inference procedure}\label{sec:infer_proc}

\begin{figure}
    \centering
    \includegraphics[width=1\columnwidth]{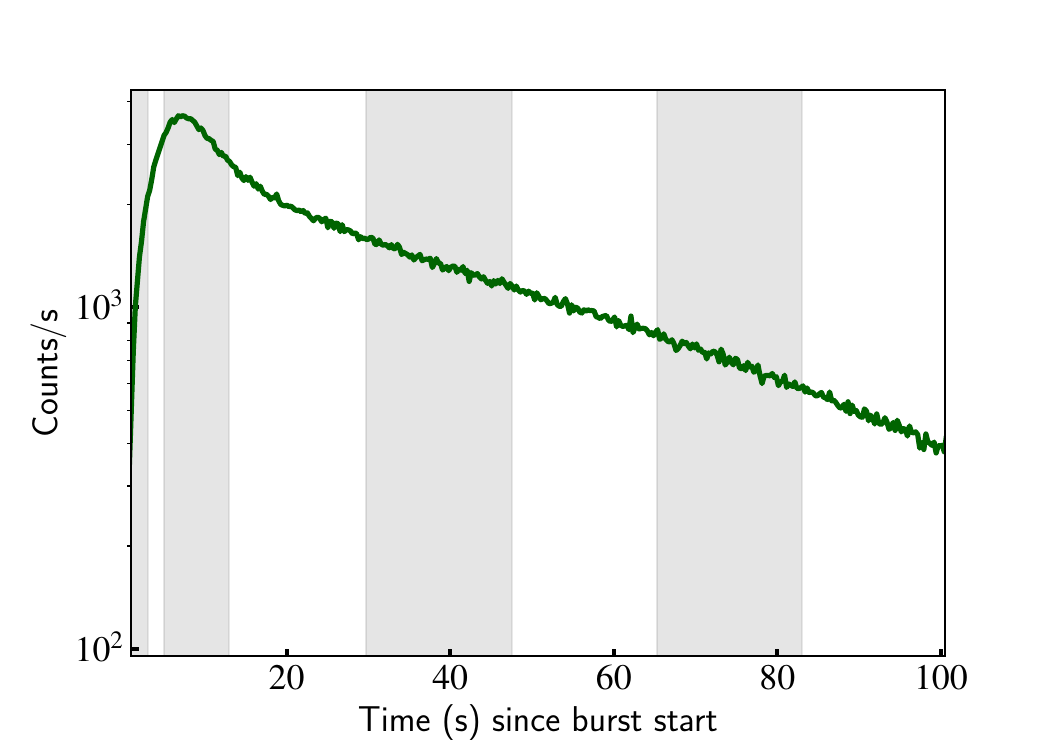}
    \caption{Example of light curve generated with parameter set 2 of model ST-$\bar{S}\tilde{H}_{T}$. The vertical bands indicate each data segment as described in the text.}
    \label{fig:burst}
    \end{figure}

To try to overcome the significant bias introduced by neglecting variability during inference as found in \citetalias{kini:2023}, we adopted a new modelling approach in this study. This method involves tracking the time-varying parameters (hot spot temperature and/or angular radius and/or star's temperature depending on the phenomenological model) to some degree during the burst. To accomplish this, we divided each burst, in each of the data sets, into eight 
time segments during which changes in flux are smaller. We elected, from visual inspection,\footnote{Another option would be to slice the bursts such that they have the same number of counts per segment or the same duration. However, to capture the rapid changes during the rise of the burst, many segments would be required, leading to a high computational cost. Increasing counts or extending the time interval would not capture rapid changes in the rise which would potentially introduce biases in mass and radius estimates.} to use two 2 s segments in the burst rise, an 8 s segment in the peak and initial decay, and then five segments of equal duration during the decay. This was intended to strike a balance between reducing computational cost (which increases with the number of segments) and minimising variation during each segment. We term this technique the slicing method, first suggested by \citet{Lo:2013ava}. An illustration is shown in Figure \ref{fig:burst}.

During the inference process, we jointly fit the data corresponding to each segment. While the common parameters (mass: $M$, equatorial radius: $R_{\rm eq}$, distance: $D$, cosine of Earth inclination to rotation axis: $\cos(i)$, phase of the hot region: $\phi_\mathrm{spot}$,  co-latitude of the centre of the hot spot: $\Theta_\mathrm{spot}$, hydrogen column density: $N_H$)\footnote{The angular radius: $\zeta_\mathrm{spot}$, the spot temperature: $T_\mathrm{spot}$ and the star  temperature: $T_\mathrm{star}$ are kept identical for each segment for models where that parameter is not time-dependent (see  Table \ref{tab:models} for a summary of the phenomenological models ).} were kept identical across each segment, the time-varying parameters (hot spot temperature and/or angular radius and/or star's temperature depending on the phenomenological model) were allowed to vary freely between segments.
The joint-fit likelihood is the product of the per-segment likelihoods (see Appendix \ref{sec:like_deriv} for derivation):
\begin{equation}\label{eq0}
\begin{split}
   \mathcal{L}_{\mathrm{total}}  =\prod_{i=1}^{K} \mathcal{L}_{i},
\end{split}
\end{equation}
where $K=8$ is the total number of segments and $\mathcal{L}_{i}$ the likelihood of the data of the $i^{\mathrm{th}}$ segment given the model.

To compute $\mathcal{L}_{\mathrm{total}} $, we used two inference methods: the brute force method (BF) and the with-model method (wM). In the BF method, only basic assumptions 
were made regarding time-dependent parameters.
We enforced a temperature hierarchy in which the temperature of each successive segment was required to be higher or lower than the preceding one in a predetermined order. These constraints were necessary as the prior space would be too large to explore otherwise, which would result in a computationally unfeasible scenario. Therefore we had:
\begin{itemize}

    \item For all phenomenological models with $T_\mathrm{X}$ varying, with $X\in \{\mathrm{spot}, \mathrm{star}\}$, we set:  $T_{X1} \le T_{X2} \le$ $T_{X3}$  and  $T_{X3} \ge T_{X4 } \ge T_{X5} \ge T_{X6} \ge$ $T_{X7} \ge$ $T_{X8}$.  
    
    \item For all phenomenological models with $\zeta_\mathrm{spot}$ varying, except for ST-$\tilde{S}_{T}\tilde{H}_{R}$, we had $\zeta_\mathrm{spot1}$ $\le$ $\zeta_\mathrm{spot2}$ $\le$ $\zeta_\mathrm{spot3}$ $\le$ $\zeta_\mathrm{spot4}$ $\le$$\zeta_\mathrm{spot5}$ $\le$$\zeta_\mathrm{spot6}$ $\le$$\zeta_\mathrm{spot7}$ $\le$$\zeta_\mathrm{spot8}$.

    \item For ST-$\tilde{S}_{T}\tilde{H}_{R}$, we had $\zeta_\mathrm{spot1}$ $\le$ $\zeta_\mathrm{spot2}$ $\le$ $\zeta_\mathrm{spot3}$ and  $\zeta_\mathrm{spot3}$  $\ge$ $\zeta_\mathrm{spot4}$ $\ge$$\zeta_\mathrm{spot5}$ $\ge$ $\zeta_\mathrm{spot6}$ $\ge$$\zeta_\mathrm{spot7}$ $\ge$$\zeta_\mathrm{spot8}$.
    
\end{itemize}
We let the simulation code \texttt{X-PSI} identify the optimal parameters that jointly fit all the data segments. 

In the second method (wM), the evolution of the time-varying parameters was assumed to be well-constrained from theory. In the interests of computational efficiency, we do this by setting the parameter prior bounds (of the hot spot temperature and/or angular radius and/or star's temperature) for a specific segment to be the minimum and the maximum of the values (of the hot spot temperature and/or angular radius and/or star's temperature) used to generate the data of that specific segment (see Table \ref{tab:Priors} in Appendix \ref{sec:appendix} for prior densities).

\subsection{Combining bursts}\label{sec:combinig_formalism}

Attempting to constrain NS properties using individual observations (bursts) results in significant uncertainties in inferred properties. Combining information from these individual bursts is necessary to overcome these limitations and achieve a more precise determination of properties commonly shared across bursts. The following section details the formalism we employed to achieve improved constraints, primarily on the mass and radius. However, this formalism can also be expanded to include other shared parameters if desired.

\subsubsection{Formalism}
We denote by $\mathcal{D}= \{d_i\}_{i=1}^{N}$ a set of data corresponding to $N$ observations (bursts) where $d_i$ is the data corresponding to the $i^{\mathrm{th}}$ burst. According to Bayes' theorem:

\begin{equation}\label{eq1}
  p(\theta|d_i)=\frac{p(d_i|\theta)p(\theta)}{p(d_i)},  
\end{equation}
where $p(\theta|d_i)$ is the posterior of the parameters $\theta$ given the data $d_i$, $p(d_i|\theta)$ the likelihood  of the data $d_i$ given the
parameters $\theta$, $p(\theta)$ the prior on $\theta$, and $p(d_i)$ the evidence.  Note that $\theta = (M, R_{\rm eq}, D, \cos(i), \phi_\mathrm{spot}, \Theta_\mathrm{spot}, \zeta_\mathrm{spot}, T_\mathrm{spot}, T_\mathrm{star}, N_H)$\footnote{When the burst is divided into K segments and the angular radius of the hot spot is kept free for each segment, then $\zeta_\mathrm{spot}=\{\zeta_\mathrm{spot1}, \zeta_\mathrm{spot2}, ..., \zeta_\mathrm{spotK}\}$\label{Zeta}. The same goes for  $T_\mathrm{spot}$ and $T_\mathrm{star}$.}. However, since the focus is on improving the constraints on the mass and radius from individual bursts, we denote by $\alpha$ a subset of $\theta, \alpha=(M, R_\mathrm {eq})$\footnote{$\theta$ can be reduced to any set of parameter of interest as long as these parameters are shared across bursts.}.  $p(\alpha|d_i)$ hence becomes the marginal posterior in the joint mass and radius space, meaning:

\begin{equation}\label{eq2}
  p(\alpha|d_i)=\int \left( \prod_{k\neq \alpha} d\theta_{k} \right)p(\theta|d_i).
\end{equation}
The main goal is to determine $p(\alpha|\mathcal{D})$ based on the $N$ bursts. Assuming these $N$ bursts are independent, and using Bayes' theorem, we have:

\begin{equation}\label{eq3}
\begin{split}
   p(\alpha|\mathcal{D})\equiv p(\alpha|\{d_i\}_{i=1}^{N}) & =\frac{p(\{d_i\}_{i=1}^{N}|\alpha)p(\alpha)}{p(\{d_i\}_{i=1}^{N})} \\
                                                           & = \dfrac{\left(\prod\limits_{i=1}^{N} p(d_i|\alpha)\right) p(\alpha)}{p(\{d_i\}_{i=1}^{N})}\\
                                                           & =\dfrac{\prod\limits_{i=1}^{N} p(\alpha|d_i) \prod\limits_{i=1}^{N}p(d_i)}{ p(\alpha)^{N-1}p(\{d_i\}_{i=1}^{N})}.\\
\end{split}
\end{equation}

Given that the joint mass-radius prior density is flat in $(M, R_{\mathrm{eq}})$ space i.e $p(\alpha)$ = constant, we obtain:

\begin{equation}\label{eq4}
\begin{split}
   p(\alpha|\mathcal{D})  \propto \prod\limits_{i=1}^{N} p(\alpha|d_i).
\end{split}
\end{equation}

Thus, the joint mass and radius posterior distribution (not normalized) over the $N$ observations can be expressed as the product of the individual posterior distributions computed for each burst.

\subsubsection{Numerical implementation}
To compute Equation (\ref{eq4}) numerically, we employed the following approach. First, we approximated the joint mass-radius posterior distribution for a specific burst $i$ by using a continuous function $f_i(\alpha)$. This function serves as an approximation for the discrete joint mass-radius posterior distribution. We obtained $f_i$ by employing a Gaussian kernel density estimator (\texttt{scipy.stats.gaussian\_kde} function) with the posterior samples from the $i^{\mathrm{th}}$ burst obtained from \texttt{MultiNest} \citep{MultiNest_2008,MultiNest_2009,MultiNest_2019} and its python wrapper \texttt{PyMultinest} \citep{pymultinest:2014}. Second, we discretized the joint mass and radius prior space into a $400\times400$ mesh grid. For each burst, we computed new weights for the mid-points of each bin within the mesh grid. These weights represent the significance of each pixel in relation to the overall posterior. Next, we multiplied the obtained weights across all bursts and subsequently normalized them. This process resulted in the desired combined posterior weights, which integrate information from all bursts. To ensure consistency with the previous analysis with \texttt{X-PSI}, we adjusted the bandwidth of the kernel density estimator based on the grid resolution. The aim was to minimize the discrepancy between the computed weights for each burst (considering the new samples taken as mid-points of the mesh grid)  and the posteriors obtained from \texttt{MultiNest} and \texttt{PyMultinest} output samples. We found that setting the bandwidth to 0.1 yields the desired result.

\section{Results}\label{sec:result}

The following section presents the results of the inference runs, which were conducted to explore parameter recovery and the associated computational costs using the slicing method. The general specifications of the hardware used for these runs are shown in  Table \ref{tab:specs} Appendix \ref{sec:appendix}. A total of 47 inference runs were performed on the first data subset, 23 using the BF method and 24 using the wM method. The results of these runs are presented in sections \ref{sec:m-r_recovery}, \ref{sec:T_R_recovery} and \ref{sec:cost}. Using the second data subset, we carried out 9 additional runs utilizing the wM approach, to examine the impact of merging posteriors from multiple bursts. Section \ref{sec:combing} shows the results of these runs.

Using the BF method, 19 runs converged, 4 were terminated before convergence due to high computing time relative to the number of accepted samples found by \texttt{MultiNest}, and one we decided not to run based on the anticipated computational cost.  This last burst was  generated using parameter set 2 of model ST-$\bar{S}\tilde{H}_{T,R}$: this star is exceedingly compact, requiring an increased number of multiple imaging evaluations in the vicinity of the true solution. This already resulted in significantly longer inference runs in \citetalias{kini:2023}, so based on the scaling of the other runs once slicing was implemented, we concluded the convergence time would be unreasonable.

\subsection{Mass and radius recovery}\label{sec:m-r_recovery}
\begin{figure}
    \centering
    \includegraphics[width=\columnwidth]{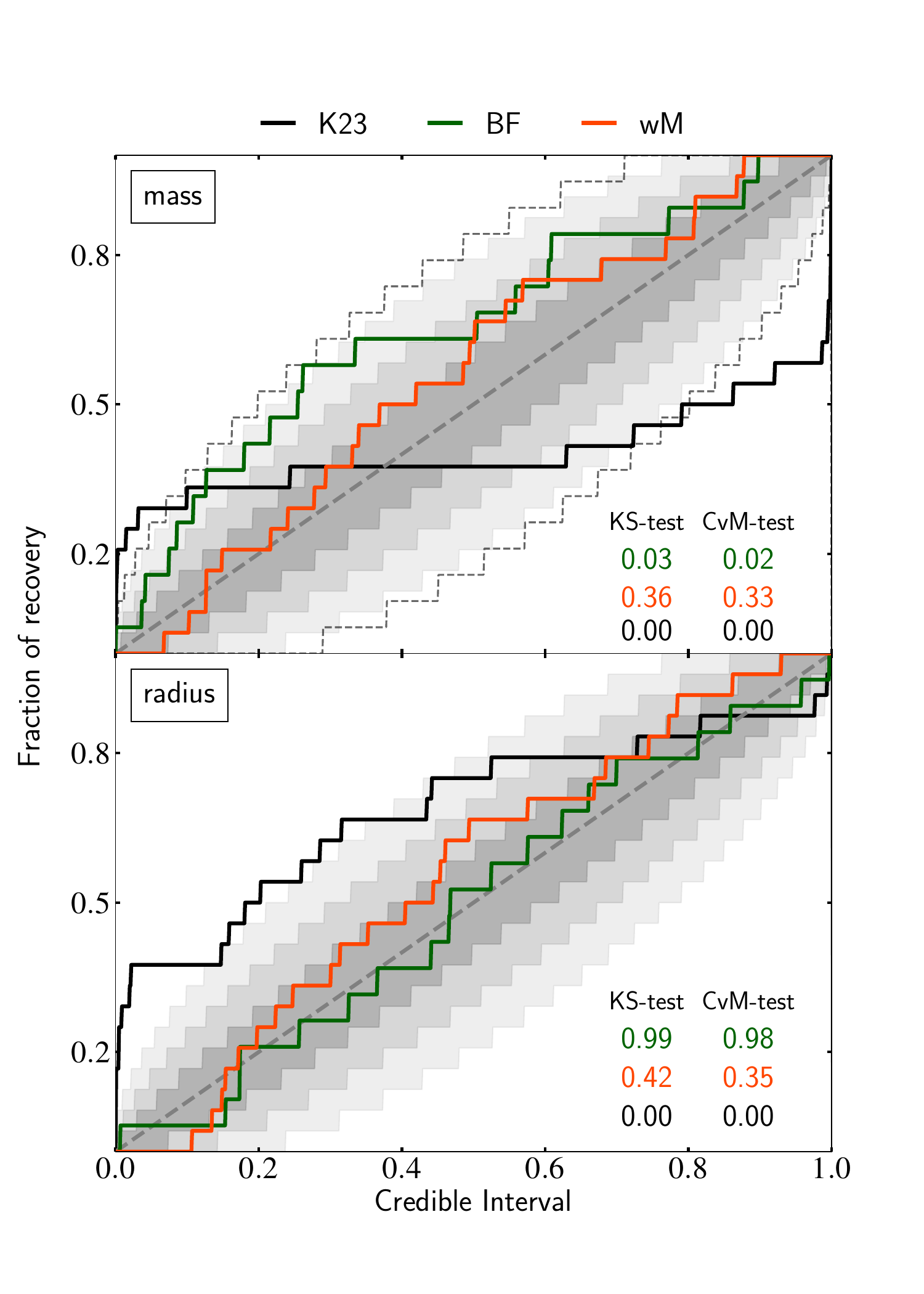}
    \caption{ Cumulative fraction of masses or radii recovered within a credible interval as a function of the credible intervals (from the lower tail of the posterior distribution, see text). The black solid line represents the result obtained in \citetalias{kini:2023}, while the green solid and orange solid lines show the results obtained using the BF method and the wM method, respectively. The color-coded values are the p-values from the KS-test and the CvM-test. The grey regions, in order of decreasing transparency, are the cumulative 1, 2, and 3$\sigma$ confidence intervals computed as in \citet{Cameron:2011}, with a sample size of 24 bursts. The upper and lower bounds of the 3$\sigma$ cumulative confidence intervals for a sample of 19 bursts, which corresponds to the completed runs using the BF method, are represented by black dashed lines in the upper panel. }
    \label{fig:pp_plot}
\end{figure}

To evaluate the mass and radius recovery, we use both PP-plots and the wrongness parameter. PP-plots provide a cumulative fraction of masses or radii recovered within a credible interval as a function of the credible intervals. We define the credible intervals for the PP-plot to always start from the  lower tail of the posterior
distribution. This is to align with the convention used by e.g. \texttt{Bilby}\footnote{\url{https://github.com/lscsoft/bilby}} \citep{Bilby}. 
On the other hand, the wrongness\footnote{For a given population X of mass or radius, wrongness $=-\frac{1}{2} z(X^{\rm{inf}})$ if $\tilde{X}=\bar{X}$. $z(X^{\rm{inf}})$, $\tilde{X}$ and $\bar{X}$ are respectively the $z\_\rm{score}$ of the inferred value $X^{\rm{inf}}$, the median and the mean of the population X.} parameter quantifies the deviation of the inferred value from the injected one, measured in units of 68\% credible intervals. A wrongness of e.g. +0.5 implies that injected value is recovered at +1$\sigma$ away from the median of the posterior distribution.
A Kolmogorov-Smirnov test (KS-test) is then conducted to examine whether the proportion of recovery conforms to a uniform distribution, which is the expected result.
The combination of these techniques was used in \citetalias{kini:2023} and provides a more comprehensive overview of parameter recovery than using either technique alone. In addition to the KS-test, we performed the Cramér–von Mises test (CvM-test) since it is more sensitive to deviations from uniformity across the entire distribution.

We show in Figure \ref{fig:pp_plot} the PP-plots for both the mass and the radius. The p-values from the KS-test and the CvM-test are indicated and colour-coded by method. Both mass and radius recovery was significantly improved by slicing and jointly fitting the data, compared to ignoring variability (solid black line). The BF method yielded slightly less accurate mass recovery than the wM method, with the mass plot consistently above the diagonal. This implies that, on average, the mass is slightly overestimated. Nevertheless, the recovered mass mostly falls within the 95\% credible region and never exceeds the 99\% credible region (corresponding to 19 bursts). In contrast, the wM model effectively recovered the mass with no observable bias, as evidenced by its p-values. While the radius in \citetalias{kini:2023} deviates significantly from the diagonal,
it is recovered here accurately (i.e. with no measurable bias) using both the BF and wM methods

\begin{figure}
    \centering
    \includegraphics[width=\columnwidth]{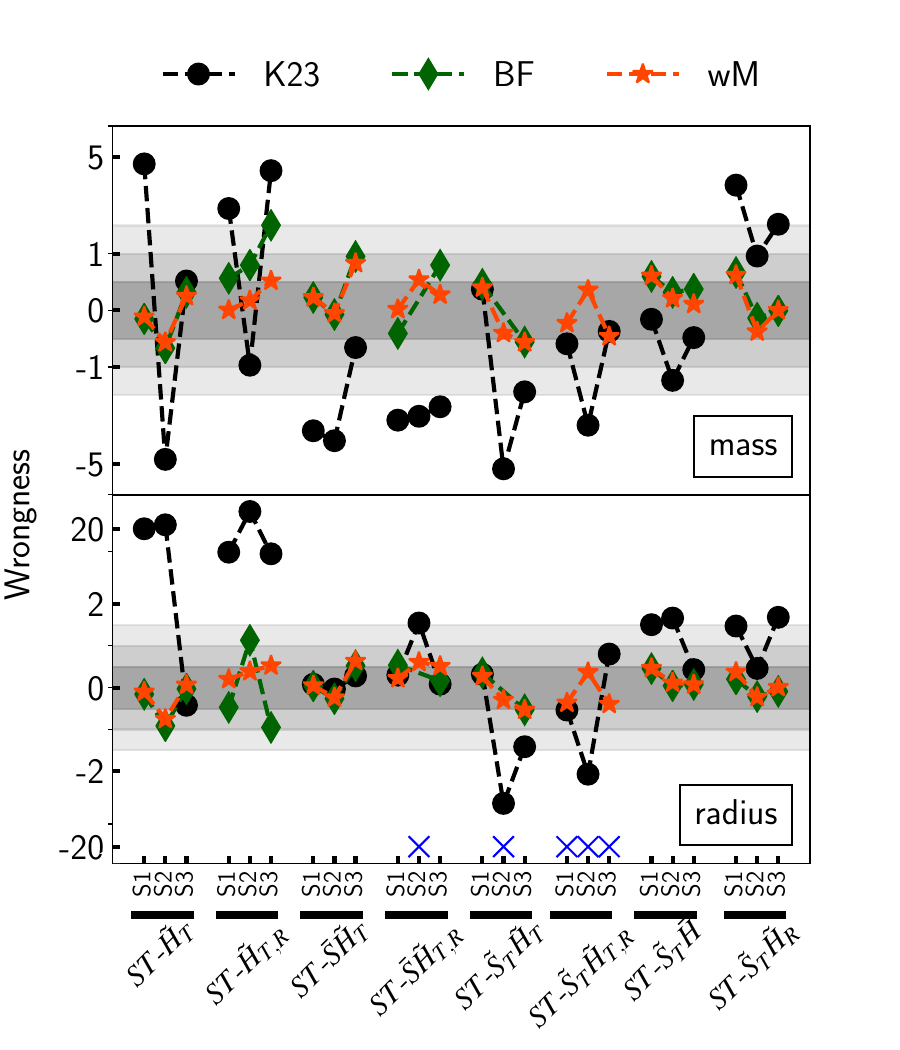}
    \caption{ Wrongness of mass and radius for both methods: slicing (BF \& wM) and \citetalias{kini:2023}. The Wrongness = $(X^{\mathrm{inf}}-X^{\mathrm{inj}})/\Delta X^{\mathrm{inf}}$, quantifies the deviation of the inferred value ($X^{\mathrm{inf}}$) from the injected
one ($X^{\mathrm{inj}})$ in units of 68\% credible interval; X $\in \{ M, R_{\mathrm{eq}}\}$. Each model is represented on the x-axis along with its corresponding parameter sets, consisting of three distinct sets: S1, S2, and S3. The grey regions, in order of decreasing transparency, are the 68\%, 95\%, and 99\% credible intervals computed as described in the text (Section \ref{sec:m-r_recovery}). The blue cross symbols show where the inference of the parameters could not be done because of computing limitations for the BF method.}
    \label{fig:wrongness}
\end{figure}

In Figure \ref{fig:wrongness}, we compare the wrongness of the parameters (mass and radius) provided by the slicing technique with those acquired without slicing as in \citetalias{kini:2023}. The blue cross symbols indicate situations in which inference was either not performed or was incomplete due to insufficient computational resources using the BF method. The grey regions, in order of decreasing transparency, are the 68\%, 95\%, and 99\% credible intervals. A simple heuristic was used to approximate the 95\% and 99\% credible interval, with these values taken to be 2 and 3 times the 68\% credible intervals respectively.
 We note that the posterior distributions are not perfectly symmetrical. As a result, a direct calculation to determine the 95\% and 99\% credible intervals from the 68\% credible intervals is not feasible without additional computational endeavour. However, they appear to be well-behaved from a visual inspection. Evaluating the wrongness for the mass and radius results generated using the slicing method results in  a clustering of wrongness around zero. This is indicative that slicing provides good parameter recovery. In fact, the majority of the mass and radius parameters were found to fall within the 68\% credible intervals, and all of them were found to be within the 99\% credible intervals. Despite the PP-plot showing that the inferred masses are mostly higher than the injected masses, from the wrongness figure, we  note that these elevated inferred masses remain mainly within the 68\% credible intervals and never exceed the pseudo 99\% credible intervals. In comparison, when variability is neglected during the inference (marked as K23 in the figure), only a limited number of parameters were determined to be within the 99\% credible region. This disparity suggests that slicing the bursts and jointly modelling the slices leads to a substantial reduction in systematic uncertainties compared to the \citetalias{kini:2023} method.

\begin{figure}
    \centering
    \includegraphics[width=\columnwidth]{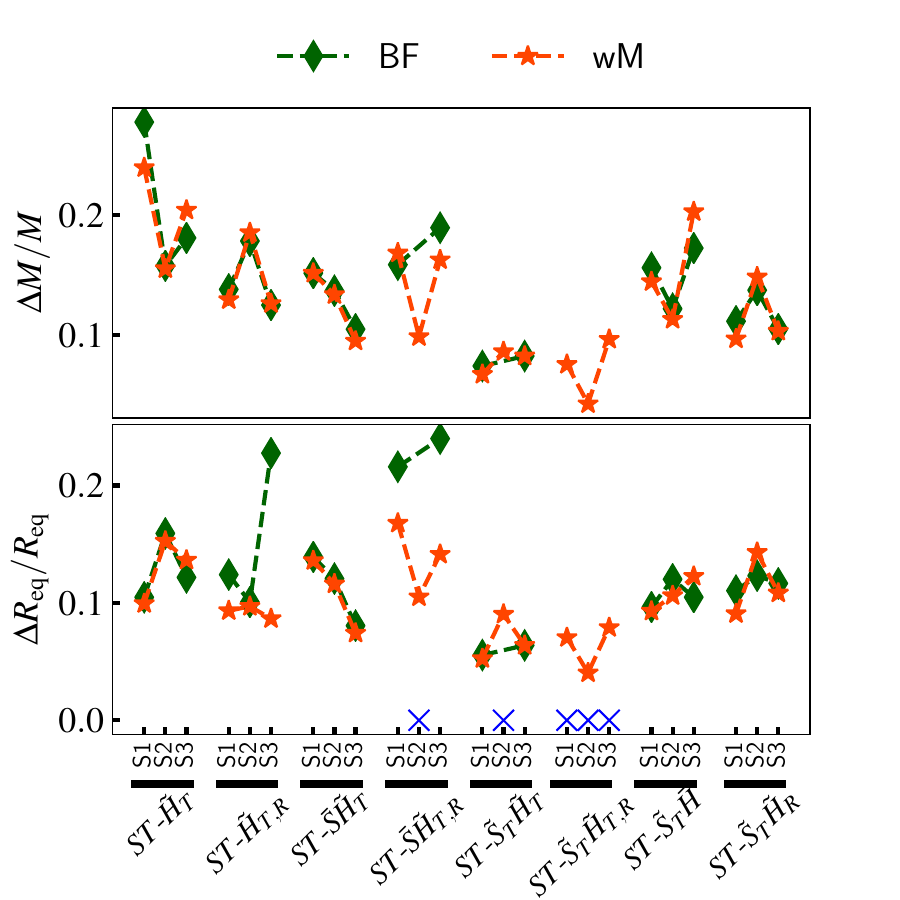}
    \caption{ Uncertainty on the mass and radius (68\% credible intervals). The x-axis shows each phenomenological model and its corresponding parameter sets, which are categorized into three distinct sets: S1, S2, and S3. The blue cross symbols indicate parameter sets where inferences were cancelled or could not be performed  due to computing limitations for the BF method.}
    \label{fig:uncertainty}
\end{figure}

Figure \ref{fig:uncertainty} shows the uncertainties at 68\% credible intervals\footnote{We denote by uncertainty at 68\% credible intervals the ratio $\Delta\mathrm{X/X^{\mathrm{inf}}}$ where $\Delta \mathrm{X}$  is the  68\% credible interval of the inferred parameter X and $X^{\mathrm{inf}}$ is the inferred median value of X.} on both the mass and radius. Typically, as the number of free parameters increases, the mass uncertainty decreases, although  there are some outliers. The decrease in the uncertainty is most probably a consequence of using a constant number of live points ($\mathrm{LP}=2000$) for all the runs. Both the BF method and the wM method  yield comparable uncertainties, with an average mass uncertainty of 12$\pm$7\% and 13$\pm$5\% when using the BF and wM methods, respectively. The mean radius uncertainty on the other hand is 10$\pm$7\% and 10$\pm$3\% when using the BF and wM methods, respectively. When not accounting for variability as in \citetalias{kini:2023}, the average uncertainty in the mass and radius was 17$\pm$7\% and 16$\pm$15\%, respectively.

\subsection{Recovered temperature and  angular radius evolution\label{sec:T_R_recovery}}

\begin{figure}
    \centering
    \includegraphics[width=\columnwidth]{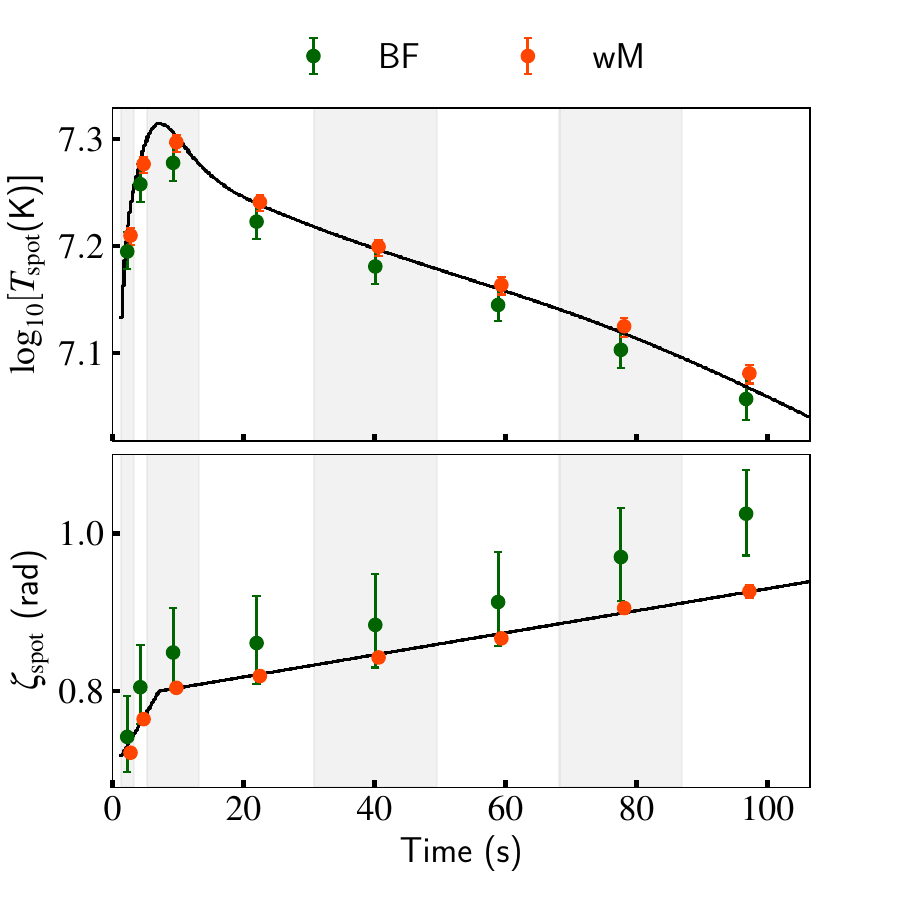}
      \caption{Example of the (median) hot spot temperatures and angular radii 
recovered using the BF and the wM methods for each segment. The error bars are the 68\% credible intervals. The injected hot spot temperature and angular radius evolutions used to generate the synthetic data are shown as solid black curves. The vertical bands, coloured in grey and white, represent the boundaries of each segment.}
    \label{fig:profile_example}
\end{figure}

In Figure \ref{fig:profile_example}, we present an illustrative example of the recovered hot spot temperature and angular radius evolution, using both the BF and wM methods. This specific example corresponds to the run performed on the data generated with model ST-$\bar{S}\tilde{H}_{T, R}$, parameter set 1. The remaining parameter evolution can be found in the Zenodo files \citep{zenodo_paper2}.

The wM method typically produces a temperature evolution that closely matches the injected evolution. In this particular example, the temperature determined for each segment corresponds to the average temperature of that segment. However, it is worth noting that this is not always the case, as there are instances where the recovered temperature may slightly exceed the segment's average \citep[see][]{zenodo_paper2}. In contrast, the angular radius of the spot is consistently equal to the average injected angular radius for all the runs with the wM method. Using the BF method, the temperatures and angular radius evolution also closely resemble the injected evolution in terms of shape. However their normalization, as compared to the injected evolutions, are not consistently equal to one.

\subsection{Combining multiple bursts}\label{sec:combing}

\begin{figure}
    \centering
    \includegraphics[width=\columnwidth]{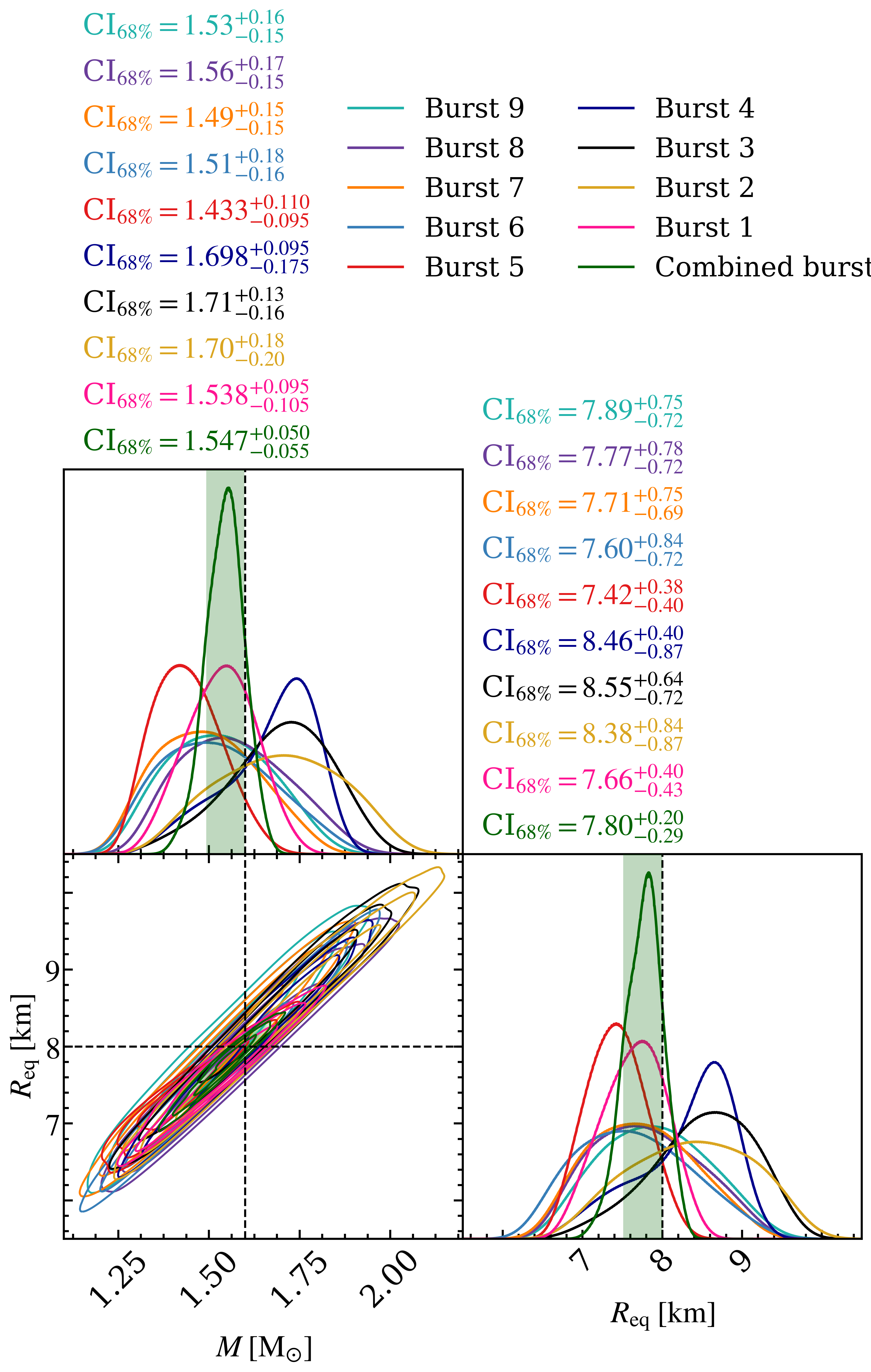}
      \caption{ Mass and radius posterior distributions for each burst and when they are combined. The dashed lines indicate the injected values ($M=1.6$ $M_{\odot}$, $R_{\rm eq}= 8.0$ km) while the green vertical bands are the 68\% credible intervals corresponding to the combined bursts. For each burst, colour-coded numbers show the median values and their corresponding 68\% credible intervals.}
    \label{fig:combing_combined}
\end{figure}

We conducted inference runs on each of the nine bursts from the second subset, with each burst containing approximately 10$^5$ counts. We used the wM method for this purpose. Figure \ref{fig:combing_combined} presents the posterior distributions for both mass and radius. Notably, the mass and radius posteriors exhibit wide dispersion around the injected values ($M=1.6$ $M_{\odot}$, $R_{\rm eq}= 8.0$ km) and vary from burst to burst. This variation can be primarily attributed to differences in burst properties, since the angular radius of the hot spot, the temperature of the hot spot, the star temperature evolutions, and the realization of noise in each simulated burst are all different. Sampling noise may also contribute to the observed differences \citep[see e.g.][]{vinciguerra:2023_sim}.

Despite the variations in the posteriors, the injected mass and radius consistently fall within the 68\% credible intervals when the wM method is applied, except for burst 5. On average, the uncertainty in mass is 18$\pm$3\%\footnote{The stated error corresponds to the standard deviation of the mass uncertainties.}, while the uncertainty in radius is 17$\pm$4\%. Combining all nine bursts using the formalism described in section \ref{sec:combing} leads to tighter constraints, as expected. The posteriors are narrower and the uncertainties on the mass and radius are respectively 7$\pm$7\%\footnote{Here, the error is derived using error propagation.} and 6$\pm$6\%. The injected radius falls within the 68\% credible interval, whereas the mass is just outside it. The wrongness of the mass is approximately -0.505 which indicates that the injected value is very close to being within the 68\% credible interval.

\subsection{Computational cost}\label{sec:cost}

\begin{figure}[h]
    \centering
    \includegraphics[width=\columnwidth]{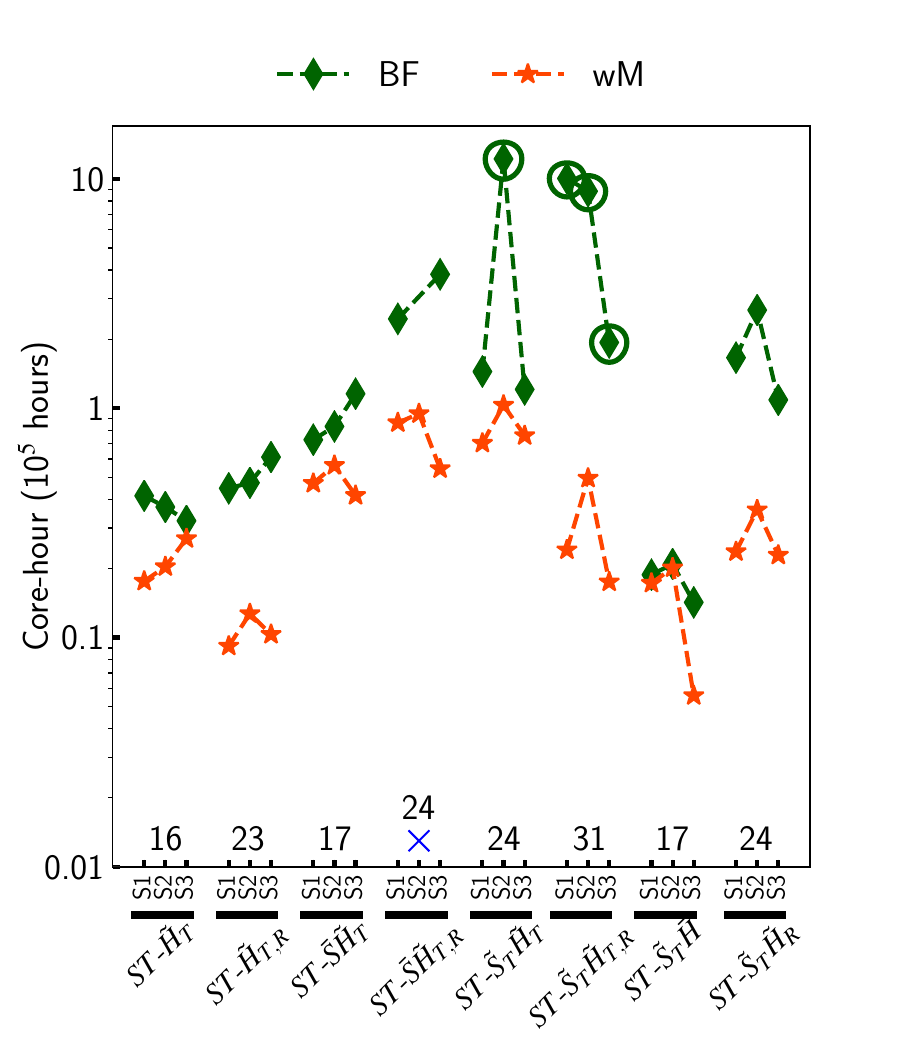}
    \caption{Core-hours spent on runs for each model and parameter set. Each model is represented on the x-axis along with its corresponding parameter sets: S1, S2, and S3. The green circles show runs that were terminated due to excessive computing demand. The blue cross indicates the parameter set where inference was not performed.}
    \label{fig:core_hours}
\end{figure}

Figure \ref{fig:core_hours} shows the computing time in core hours for each run. The blue cross at the bottom represents model ST-$\bar{S}\tilde{H}_{T,R}$ parameter set 2, where the inference run was not conducted. The number of free parameters in each model is indicated by the numbers at the bottom, and sets highlighted by a green circle are those that had their inference interrupted before convergence. The full set of all runs cost approximately 5$\times$10$^6$ core-hours in total. Typically, a single inference run takes on average 10$^5$ core-hours, and models with more free parameters take longer to converge. An exception is ST-$\bar{S}\tilde{H}_{T}$, which has 17 parameters but took as long as the models with 24 parameters.

The ST-$\tilde{S}_{T}\tilde{H}_{T,R}$ runs, along with the ST-$\tilde{S}_{T}\tilde{H}_{T}$-set 2 run, were stopped since the number of core-hours spent was extremely high. The ST-$\tilde{S}_{T}\tilde{H}_{T,R}$ simulations in particular consumed in total approximately 1.5 million core-hours without converging.
Unquestionably, the large number of free parameters in the model contributed to this substantial use of computation resources.  ST-$\tilde{S}_{T}\tilde{H}_{T}$-set 2 and ST-$\bar{S}\tilde{H}_{T,R}$-set 2 both assume an extremely compact star. This leads to the computation of higher images, which slows down the likelihood evaluation as the sampler homes in on the injected solution, leading to substantial consumption of computation resources.

\section{Discussion}\label{sec:discussion}

We conducted this analysis with three primary objectives in view. The first objective was to address the systematic bias that arises when TBO source properties are estimated using PPM without accounting for temporal variability, identified in \citetalias{kini:2023} as being particularly problematic once burst counts exceed $\sim$10$^6$. Secondly, we aimed to assess the computational resources needed to address these biases, to help plan the allocation of computing resources for analyses of current and future TBO data. Finally, we aimed to explore how to combine the results of individual bursts to yield more robust results. This will help us to exploit existing TBO data captured by RXTE.

Our study used synthetic burst data generated assuming various phenomenological models for the time evolution of burst and hot spot.  We first looked at slicing bursts into shorter segments, during which variability can be neglected, and jointly fitting those segments. To carry out the inference runs, we used two different methods. In one (the BF method) only very broad prior knowledge of the time-varying parameters was assumed, leaving the inference code \texttt{X-PSI} to identify the time variation. In the other (wM) we assumed some knowledge of the time-varying parameters was available from theory.  Finally, we looked at combining individual bursts whose properties, even from a single source, may differ.

\subsection{Constraints on the mass and radius}

The BF method and the wM method both lead to accurate recovery of the mass and radius, with negligible differences between the radius estimations. Almost all of the recovered radii fall within the 68\% credible intervals. While the inferred masses are generally slightly larger than the injected masses, the majority fall within the 68\% credible intervals, and all injected masses fall within the pseudo 99\% credible intervals, as indicated by the wrongness plots. Both techniques yield mass and radius estimates with approximately 10\% uncertainty (68\% credible interval) for bursts with 10$^6$ counts.  This result highlights the significant potential of TBO sources in constraining the EOS, since 5\%-10\% uncertainty is necessary for distinguishing between different EOS \citep[see e.g.][]{Ozel:2009b,Psaltis:2013fha}.

Slicing bursts containing 10$^5$ counts and performing joint fitting for each segment resulted in a decrease in uncertainty from 30\% (as found in \citetalias{kini:2023}) to approximately 20\%. Subsequently combining the posterior distributions of nine bursts, totalling around 10$^6$ counts, yielded uncertainties of about 7$\pm$7\% and 6$\pm$6\% for mass and radius measurements, respectively. These findings align closely with the $\sim 10$\% uncertainty observed when using the slicing method on a single burst of approximately 10$^6$ counts for both parameters. This implies that merging results from individual bursts can improve the constraints, highlighting the significant potential of TBO sources, especially considering that sources typically exhibit multiple bursts during their outburst phase. With next-generation instruments like eXTP and STROBE-X, which would have larger effective areas and will capture more photons, we anticipate that capturing a few bursts will significantly reduce uncertainties in the mass-radius space. Further reduction could be possible if we had independent prior constraints e.g., on the mass and inclination \citep[see e.g.][although the constraints in that specific paper are rather broad]{Wang:2017}.\footnote{Despite not making any prior assumption on the mass and inclination, the result of this analysis is not a worst-case scenario, because we have made some prior assumptions on e.g. distance and background. However nor is it necessarily a best-case scenario (J1814, for example, has 28 rather than only 10 bursts).}

\subsection{ Recovered temperature and  angular radius evolutions}
Determining, from first principles, the temperature and angular radius that accurately represent the observed counts for each segment is challenging due to numerous factors, such as atmosphere reprocessing. However, by employing the wM method and constraining the temperature and/or angular radius of each segment within the range of the lowest and highest limits used in generating the data, we mostly retrieve the average temperature. This is not always the case with the BF method, which may seem a little surprising since it shows strong agreement in matching the masses and radii obtained through the wM method, which does recover the average temperature. This is because of the multiple degeneracies between parameters, especially between the spot properties and the distance.

For the BF method, it could be argued that imposing some constraints (see section \ref{sec:infer_proc}) on the behaviour of temperature and radius evolutions might have contributed to the accurate recovery of the shape of the temperature and angular radius evolutions. However, it is important to note that this approach allows for a wide range of other possible temperature and radius evolutions. In the case of a real burst, similar constraints could reasonably be applied to the temperature evolution as the temperature curve is expected to exhibit similar behaviour to the light curve during a burst. However, the situation is different for the angular radius evolution, as the geometry of the hot spot during a burst remains poorly understood.

\subsection{Combining bursts}

As shown in section \ref{sec:combing}, posteriors obtained from multiple bursts from the same source can potentially offer more precise and reliable constraints on stellar properties. When combining multiple bursts, the posteriors of mass and radius become more tightly constrained, with uncertainties reducing from approximately 20\% to about 6-7\% compared to a single burst. This posterior shrinkage aligns with the expected $\sqrt{N}$ (where $N$ is the number of bursts) improvement compared to that of a single observation.  This is encouraging for attempts to obtain PPM constraints from TBOs in the RXTE archive.

However, it is important to consider that while combining multiple bursts yields similar outcomes to a single burst with equivalent counts, modelling a single burst with higher counts can reveal peculiarities in the model that may not be apparent in individual bursts with lower counts. Therefore, continually adding posteriors from low-count bursts to obtain better constraints may result in flawed results, especially if the total counts reach a range where statistical uncertainties become lower than systematic uncertainties.  Unless a diagnostic method ensures that the combined posteriors behave as expected, caution should be exercised when combining too many bursts.

Future data from proposed large-area X-ray spectral-timing telescopes like eXTP and STROBE-X will undoubtedly contribute to reducing uncertainties in the employed models (surface patterns, atmospheres, etc) given the data quality expected.  With these missions, we expect to be able to obtain interesting constraints even from a single burst. However the strongest constraints will come from combining multiple bursts, providing that model uncertainties can indeed be reduced. If bursts from the same star were captured with RXTE and also by newer instruments like XTP, and STROBE-X—they can, in principle, be combined to yield improved uncertainties. However, the newer, larger instruments would statistically dominate the results, given their better data quality.

\subsection{Computational cost}
Although the BF and wM methods recover mass and radius with comparable accuracy, the computational time required for each method differs significantly. On average, it took about $4\times10^4$ core hours for the runs to converge using the wM method, while the BF method required approximately 10$^5$ core hours. Runs conducted on the identical dataset without employing data slicing, by contrast, took on average $10^4$ core hours\footnote{Note however that these runs used much higher energy resolution. See the discussion that follows.} \citepalias{kini:2023}. As expected, phenomenological models with more free parameters generally took longer to converge. However, the completed runs were still three times faster when using the wM method compared to the BF method. Although more than $3\times10^6$ core hours were spent on set 2 of ST-$\tilde{S}_{T}\tilde{H}_{T}$ and all  the sets of ST-ST-$\tilde{S}_{T}\tilde{H}_{T,R}$ using the BF method, those runs failed to converge. This was primarily due to the extensive prior space that needed to be explored for some models, often combined with the very compact star assumption used to generate these synthetic data.

The computational demands of the slicing method stem from the need to compute a pulse profile for each segment. This involves performing tasks like ray tracing and interpolating atmospheric properties for each segment. In an attempt to expedite the computations, we significantly reduced the resolution in the energy grid used for pulse profile computation compared to that used in \citetalias{kini:2023}\footnote{The synthetic data was created using an energy grid resolution of 128. In \citetalias{kini:2023}, the same grid resolution was used during inference while in this analysis, we used 16.}. However, the computational cost remains substantial.

When modelling sources with higher counts, it may be necessary to include even more time segments to effectively address mass and radius biases. This would result in further increases in computation time. Computational costs could also escalate if other complexities need to be taken into account. Even in the case of J1814, considered to be one of the most static sources in terms of temporal variability during the burst, the hot spot location is observed to change slightly \citep{Watts:2008,Cavecchi:2022}. Investigating how such complexity affects mass and radius estimation is still an open question. For many sources \citep[see e.g.][]{Strohmayer:1996,muno02a,Muno:2002,Chakrabarty:2003,Bhattacharyya:2005, Bhattacharyya:2006, Bilous:2019} the changes in some of the pulsation properties are more pronounced.  Ideally, keeping the parameters describing the hot spot properties as well as the stellar temperature as free parameters between segments would be desirable.

    How might computational cost be reduced? The wM method highlights that despite the inherent complexity, theoretical work to establish expected behaviour for these variabilities can significantly reduce computational time. In addition, \texttt{X-PSI} or any other pulse profile modelling codes, could be more optimized for analyses with many time segments.
    We could also explore other sampling options, as current nested samplers are known to require many proposals for large parameter spaces to be  explored thoroughly. Recent advances in machine learning techniques may offer a worthwhile alternative \citep[see e.g.][]{Baron:2019, Chua:2020, Bhardwaj:2023}. Using Surrogate models (neural network emulators) could potentially help create a faster simulation model.

\subsection{Limitations and futures prospects }

This study, just like \citetalias{kini:2023}, is limited to synthetic data sets that have been tailored to mimic the properties of one particularly promising source, J1814.  And due to the computational cost of the slicing method, we have in this paper conducted parameter recovery only using the $10^6$ counts data set from \citetalias{kini:2023}.   
We chose not to include the $10^7$ counts data in our analysis for two main reasons: the anticipated (even higher) computational cost of a larger data set; and the fact that  
the total number of counts for the J1814 bursts recorded by RXTE during its 2003 outburst, which we plan to analyse in a future paper, is a few times $10^6$. We also considered only one particular slicing formulation (with eight segments). Within these limitations, however, both slicing and combining bursts improves parameter recovery to the desired level. 

Ultimately, we would like to perform PPM on a wider variety of TBO sources (with much more variability than J1814, including hot spot motion), for larger data sets with $\sim 10^7$ counts, and using different instrument response matrices.  While prospects for the slicing method, and for combining bursts, look promising, verifying that this holds for the more general case - and optimizing the technique to minimize computational cost - will certainly require additional exploratory simulations.

\section{Conclusion}\label{sec:Conclusion}

In this series of papers (\citetalias{kini:2023} and the current one), we have undertaken an exploration of the challenges and opportunities associated with the use of PPM to deduce the masses and radii of neutron stars. Our specific focus has been on Thermonuclear Burst Oscillation (TBO) sources.

We have found that slicing bursts into shorter segments, where temporal variability can be neglected, allows good parameter recovery for count rates at the level necessary to place meaningful constraints on the EoS.  
This process of slicing the data does however come with a notable computational cost. Thus, to accurately model data from bursting sources, especially those obtained with next-generation instruments, it will be imperative to allocate sufficient computational resources. The development of theoretical models that better constrain the properties of bursts and oscillations would also be of significant assistance.

We also explored strategies for combining information obtained from multiple individual bursts to achieve more precise constraints for a given source. By merging bursts, we can overcome the limitations of individual observations and enhance the statistical significance of our findings.

TBO sources hold great promise for contributing to a deeper comprehension of the core composition of neutron stars and providing valuable insights into the fundamental properties of matter under extreme conditions. Moving forward, further advancements in observational capabilities, coupled with ongoing theoretical developments, will continually enhance our understanding of the physics and astrophysics of neutron stars.

\section*{Acknowledgements}

YK, TS, SV, ALW and DC acknowledge support from ERC Consolidator Grant No.~865768 AEONS (PI Watts). The major part of the work was carried out on the HELIOS cluster exclusively on dedicated nodes funded via the abovementioned ERC CoG. We thank SURF (\url{www.surf.nl}) for the support in using the Lisa Compute Cluster. We also thank Phil Uttley for discussion on rms FA. ZM was supported by the U.S. Department of Energy Office of Science under Grants No. DE-FG02-88ER40387 and DE- SC0019042, the U.S. National Nuclear Security Administration through Grant No. DE-NA0003909, and the U.S. National Science Foundation under Grant No. OISE-1927130 (International Research Network for Nuclear Astrophysics). VS acknowledges support by Deutsche Forschungsgemeinschaft (DFG; grant WE 1312/59-1).

\section*{Data availability}
We used the simulation code X-PSI publicly available at \url{https://github.com/xpsi-group/xpsi.git}. Synthetic data, posterior samples  and post-processing notebooks are available on Zenodo (see: \url{ https://doi.org/10.5281/zenodo.8033527})



\bibliographystyle{mnras}
\bibliography{bibliography}

\begin{thebibliography}{}
\makeatletter
\relax
\def\mn@urlcharsother{\let\do\@makeother \do\$\do\&\do\#\do\^\do\_\do\%\do\~}
\def\mn@doi{\begingroup\mn@urlcharsother \@ifnextchar [ {\mn@doi@}
  {\mn@doi@[]}}
\def\mn@doi@[#1]#2{\def\@tempa{#1}\ifx\@tempa\@empty \href
  {http://dx.doi.org/#2} {doi:#2}\else \href {http://dx.doi.org/#2} {#1}\fi
  \endgroup}
\def\mn@eprint#1#2{\mn@eprint@#1:#2::\@nil}
\def\mn@eprint@arXiv#1{\href {http://arxiv.org/abs/#1} {{\tt arXiv:#1}}}
\def\mn@eprint@dblp#1{\href {http://dblp.uni-trier.de/rec/bibtex/#1.xml}
  {dblp:#1}}
\def\mn@eprint@#1:#2:#3:#4\@nil{\def\@tempa {#1}\def\@tempb {#2}\def\@tempc
  {#3}\ifx \@tempc \@empty \let \@tempc \@tempb \let \@tempb \@tempa \fi \ifx
  \@tempb \@empty \def\@tempb {arXiv}\fi \@ifundefined
  {mn@eprint@\@tempb}{\@tempb:\@tempc}{\expandafter \expandafter \csname
  mn@eprint@\@tempb\endcsname \expandafter{\@tempc}}}

\bibitem[\protect\citeauthoryear{{AlGendy} \& {Morsink}}{{AlGendy} \&
  {Morsink}}{2014}]{AlGendy:2014}
{AlGendy} M.,  {Morsink} S.~M.,  2014, \mn@doi [\apj]
  {10.1088/0004-637X/791/2/78}, \href
  {https://ui.adsabs.harvard.edu/abs/2014ApJ...791...78A} {791, 78}

\bibitem[\protect\citeauthoryear{{Ashton} et~al.,}{{Ashton}
  et~al.}{2019}]{Bilby}
{Ashton} G.,  et~al., 2019, \mn@doi [\apjs] {10.3847/1538-4365/ab06fc}, \href
  {https://ui.adsabs.harvard.edu/abs/2019ApJS..241...27A} {241, 27}

\bibitem[\protect\citeauthoryear{{Baron}}{{Baron}}{2019}]{Baron:2019}
{Baron} D.,  2019, \mn@doi [arXiv e-prints] {10.48550/arXiv.1904.07248}, \href
  {https://ui.adsabs.harvard.edu/abs/2019arXiv190407248B} {p. arXiv:1904.07248}

\bibitem[\protect\citeauthoryear{{Baub{\"o}ck}, {Psaltis}, {{\"O}zel}  \&
  {Johannsen}}{{Baub{\"o}ck} et~al.}{2012}]{Baubock:2012}
{Baub{\"o}ck} M.,  {Psaltis} D.,  {{\"O}zel} F.,   {Johannsen} T.,  2012,
  \mn@doi [\apj] {10.1088/0004-637X/753/2/175}, \href
  {https://ui.adsabs.harvard.edu/abs/2012ApJ...753..175B} {753, 175}

\bibitem[\protect\citeauthoryear{Baym, Hatsuda, Kojo, Powell, Song  \&
  Takatsuka}{Baym et~al.}{2018}]{Baym:2017}
Baym G.,  Hatsuda T.,  Kojo T.,  Powell P.~D.,  Song Y.,   Takatsuka T.,  2018,
  \mn@doi [Rept. Prog. Phys.] {10.1088/1361-6633/aaae14}, 81, 056902

\bibitem[\protect\citeauthoryear{{Beloborodov}}{{Beloborodov}}{2002}]{Beloborodov:2002}
{Beloborodov} A.~M.,  2002, \mn@doi [\apjl] {10.1086/339511}, \href
  {https://ui.adsabs.harvard.edu/abs/2002ApJ...566L..85B} {566, L85}

\bibitem[\protect\citeauthoryear{{Bhardwaj}, {Alvey}, {Miller}, {Nissanke}  \&
  {Weniger}}{{Bhardwaj} et~al.}{2023}]{Bhardwaj:2023}
{Bhardwaj} U.,  {Alvey} J.,  {Miller} B.~K.,  {Nissanke} S.,   {Weniger} C.,
  2023, \mn@doi [arXiv e-prints] {10.48550/arXiv.2304.02035}, \href
  {https://ui.adsabs.harvard.edu/abs/2023arXiv230402035B} {p. arXiv:2304.02035}

\bibitem[\protect\citeauthoryear{{Bhattacharyya} \&
  {Strohmayer}}{{Bhattacharyya} \& {Strohmayer}}{2005}]{Bhattacharyya:2005}
{Bhattacharyya} S.,  {Strohmayer} T.~E.,  2005, \mn@doi [\apjl]
  {10.1086/499100}, \href
  {https://ui.adsabs.harvard.edu/abs/2005ApJ...634L.157B} {634, L157}

\bibitem[\protect\citeauthoryear{{Bhattacharyya} \&
  {Strohmayer}}{{Bhattacharyya} \& {Strohmayer}}{2006}]{Bhattacharyya:2006}
{Bhattacharyya} S.,  {Strohmayer} T.~E.,  2006, \mn@doi [\apjl]
  {10.1086/504841}, \href
  {https://ui.adsabs.harvard.edu/abs/2006ApJ...642L.161B} {642, L161}

\bibitem[\protect\citeauthoryear{Bhattacharyya, Strohmayer, Miller  \&
  Markwardt}{Bhattacharyya et~al.}{2005}]{Bhattacharyya:2004pp}
Bhattacharyya S.,  Strohmayer T.~E.,  Miller M.~C.,   Markwardt C.~B.,  2005,
  \mn@doi [\apj] {10.1086/426383}, 619, 483

\bibitem[\protect\citeauthoryear{{Bilous} \& {Watts}}{{Bilous} \&
  {Watts}}{2019}]{Bilous:2019}
{Bilous} A.~V.,  {Watts} A.~L.,  2019, \mn@doi [\apjs]
  {10.3847/1538-4365/ab2fe1}, \href
  {https://ui.adsabs.harvard.edu/abs/2019ApJS..245...19B} {245, 19}

\bibitem[\protect\citeauthoryear{{Bogdanov} et~al.,}{{Bogdanov}
  et~al.}{2019a}]{Bogdanov:2019a}
{Bogdanov} S.,  et~al., 2019a, \mn@doi [\apjl] {10.3847/2041-8213/ab53eb},
  \href {https://ui.adsabs.harvard.edu/abs/2019ApJ...887L..25B} {887, L25}

\bibitem[\protect\citeauthoryear{{Bogdanov} et~al.,}{{Bogdanov}
  et~al.}{2019b}]{Bogdanov:2019}
{Bogdanov} S.,  et~al., 2019b, \mn@doi [\apjl] {10.3847/2041-8213/ab5968},
  \href {https://ui.adsabs.harvard.edu/abs/2019ApJ...887L..26B} {887, L26}

\bibitem[\protect\citeauthoryear{{Braje}, {Romani}  \& {Rauch}}{{Braje}
  et~al.}{2000}]{Braje:2000}
{Braje} T.~M.,  {Romani} R.~W.,   {Rauch} K.~P.,  2000, \mn@doi [\apj]
  {10.1086/308448}, \href
  {https://ui.adsabs.harvard.edu/abs/2000ApJ...531..447B} {531, 447}

\bibitem[\protect\citeauthoryear{{Buchner} et~al.,}{{Buchner}
  et~al.}{2014}]{pymultinest:2014}
{Buchner} J.,  et~al., 2014, \mn@doi [\aap] {10.1051/0004-6361/201322971},
  \href {https://ui.adsabs.harvard.edu/abs/2014A&A...564A.125B} {564, A125}

\bibitem[\protect\citeauthoryear{{Cadeau}, {Morsink}, {Leahy}  \&
  {Campbell}}{{Cadeau} et~al.}{2007}]{Cadeau:2007}
{Cadeau} C.,  {Morsink} S.~M.,  {Leahy} D.,   {Campbell} S.~S.,  2007, \mn@doi
  [\apj] {10.1086/509103}, \href
  {https://ui.adsabs.harvard.edu/abs/2007ApJ...654..458C} {654, 458}

\bibitem[\protect\citeauthoryear{{Cameron}}{{Cameron}}{2011}]{Cameron:2011}
{Cameron} E.,  2011, \mn@doi [\pasa] {10.1071/AS10046}, \href
  {https://ui.adsabs.harvard.edu/abs/2011PASA...28..128C} {28, 128}

\bibitem[\protect\citeauthoryear{{Cavecchi} \& {Patruno}}{{Cavecchi} \&
  {Patruno}}{2022}]{Cavecchi:2022}
{Cavecchi} Y.,  {Patruno} A.,  2022, \mn@doi [\mnras] {10.1093/mnras/stab3536},
  \href {https://ui.adsabs.harvard.edu/abs/2022MNRAS.510.1431C} {510, 1431}

\bibitem[\protect\citeauthoryear{{Cavecchi}, {Watts}, {Braithwaite}  \&
  {Levin}}{{Cavecchi} et~al.}{2013}]{Cavecchi:2013}
{Cavecchi} Y.,  {Watts} A.~L.,  {Braithwaite} J.,   {Levin} Y.,  2013, \mn@doi
  [\mnras] {10.1093/mnras/stt1273}, \href
  {https://ui.adsabs.harvard.edu/abs/2013MNRAS.434.3526C} {434, 3526}

\bibitem[\protect\citeauthoryear{{Cavecchi}, {Watts}, {Levin}  \&
  {Braithwaite}}{{Cavecchi} et~al.}{2015}]{Cavecchi:2015}
{Cavecchi} Y.,  {Watts} A.~L.,  {Levin} Y.,   {Braithwaite} J.,  2015, \mn@doi
  [\mnras] {10.1093/mnras/stu2764}, \href
  {https://ui.adsabs.harvard.edu/abs/2015MNRAS.448..445C} {448, 445}

\bibitem[\protect\citeauthoryear{{Cavecchi}, {Levin}, {Watts}  \&
  {Braithwaite}}{{Cavecchi} et~al.}{2016}]{Cavecchi:2016}
{Cavecchi} Y.,  {Levin} Y.,  {Watts} A.~L.,   {Braithwaite} J.,  2016, \mn@doi
  [\mnras] {10.1093/mnras/stw728}, \href
  {https://ui.adsabs.harvard.edu/abs/2016MNRAS.459.1259C} {459, 1259}

\bibitem[\protect\citeauthoryear{{Chakrabarty}, {Morgan}, {Muno}, {Galloway},
  {Wijnands}, {van der Klis}  \& {Markwardt}}{{Chakrabarty}
  et~al.}{2003}]{Chakrabarty:2003}
{Chakrabarty} D.,  {Morgan} E.~H.,  {Muno} M.~P.,  {Galloway} D.~K.,
  {Wijnands} R.,  {van der Klis} M.,   {Markwardt} C.~B.,  2003, \mn@doi [\nat]
  {10.1038/nature01732}, \href
  {https://ui.adsabs.harvard.edu/abs/2003Natur.424...42C} {424, 42}

\bibitem[\protect\citeauthoryear{{Chambers} \& {Watts}}{{Chambers} \&
  {Watts}}{2020}]{Chambers:2020}
{Chambers} F.~R.~N.,  {Watts} A.~L.,  2020, \mn@doi [\mnras]
  {10.1093/mnras/stz3449}, \href
  {https://ui.adsabs.harvard.edu/abs/2020MNRAS.491.6032C} {491, 6032}

\bibitem[\protect\citeauthoryear{{Chambers}, {Watts}, {Keek}, {Cavecchi}  \&
  {Garcia}}{{Chambers} et~al.}{2019}]{Chambers:2019}
{Chambers} F. R.~N.,  {Watts} A.~L.,  {Keek} L.,  {Cavecchi} Y.,   {Garcia} F.,
   2019, \mn@doi [\apj] {10.3847/1538-4357/aaf501}, \href
  {https://ui.adsabs.harvard.edu/abs/2019ApJ...871...61C} {871, 61}

\bibitem[\protect\citeauthoryear{{Chen} \& {Shaham}}{{Chen} \&
  {Shaham}}{1989}]{Chen:1989}
{Chen} K.,  {Shaham} J.,  1989, \mn@doi [\apj] {10.1086/167295}, \href
  {https://ui.adsabs.harvard.edu/abs/1989ApJ...339..279C} {339, 279}

\bibitem[\protect\citeauthoryear{{Chua} \& {Vallisneri}}{{Chua} \&
  {Vallisneri}}{2020}]{Chua:2020}
{Chua} A. J.~K.,  {Vallisneri} M.,  2020, \mn@doi [\prl]
  {10.1103/PhysRevLett.124.041102}, \href
  {https://ui.adsabs.harvard.edu/abs/2020PhRvL.124d1102C} {124, 041102}

\bibitem[\protect\citeauthoryear{{Degenaar} \& {Suleimanov}}{{Degenaar} \&
  {Suleimanov}}{2018}]{Degenaar:2018}
{Degenaar} N.,  {Suleimanov} V.~F.,  2018, \mn@doi [arXiv e-prints]
  {10.48550/arXiv.1806.02833}, \href
  {https://ui.adsabs.harvard.edu/abs/2018arXiv180602833D} {p. arXiv:1806.02833}

\bibitem[\protect\citeauthoryear{{Edelson}, {Krolik}  \& {Pike}}{{Edelson}
  et~al.}{1990}]{Edelson}
{Edelson} R.~A.,  {Krolik} J.~H.,   {Pike} G.~F.,  1990, \mn@doi [\apj]
  {10.1086/169036}, \href
  {https://ui.adsabs.harvard.edu/abs/1990ApJ...359...86E} {359, 86}

\bibitem[\protect\citeauthoryear{{Feroz} \& {Hobson}}{{Feroz} \&
  {Hobson}}{2008}]{MultiNest_2008}
{Feroz} F.,  {Hobson} M.~P.,  2008, \mn@doi [\mnras]
  {10.1111/j.1365-2966.2007.12353.x}, \href
  {https://ui.adsabs.harvard.edu/abs/2008MNRAS.384..449F} {384, 449}

\bibitem[\protect\citeauthoryear{{Feroz}, {Hobson}  \& {Bridges}}{{Feroz}
  et~al.}{2009}]{MultiNest_2009}
{Feroz} F.,  {Hobson} M.~P.,   {Bridges} M.,  2009, \mn@doi [\mnras]
  {10.1111/j.1365-2966.2009.14548.x}, \href
  {http://adsabs.harvard.edu/abs/2009MNRAS.398.1601F} {398, 1601}

\bibitem[\protect\citeauthoryear{{Feroz}, {Hobson}, {Cameron}  \&
  {Pettitt}}{{Feroz} et~al.}{2019}]{MultiNest_2019}
{Feroz} F.,  {Hobson} M.~P.,  {Cameron} E.,   {Pettitt} A.~N.,  2019, \mn@doi
  [The Open Journal of Astrophysics] {10.21105/astro.1306.2144}, \href
  {https://ui.adsabs.harvard.edu/abs/2019OJAp....2E..10F} {2, 10}

\bibitem[\protect\citeauthoryear{{Galloway} et~al.,}{{Galloway}
  et~al.}{2020}]{Galloway:2020}
{Galloway} D.~K.,  et~al., 2020, \mn@doi [\apjs] {10.3847/1538-4365/ab9f2e},
  \href {https://ui.adsabs.harvard.edu/abs/2020ApJS..249...32G} {249, 32}

\bibitem[\protect\citeauthoryear{{Garcia}, {Chambers}  \& {Watts}}{{Garcia}
  et~al.}{2018}]{Garcia:2018}
{Garcia} F.,  {Chambers} F.~R.~N.,   {Watts} A.~L.,  2018, \mn@doi [Physical
  Review Fluids] {10.1103/PhysRevFluids.3.123501}, \href
  {https://ui.adsabs.harvard.edu/abs/2018PhRvF...3l3501G} {3, 123501}

\bibitem[\protect\citeauthoryear{{Gendreau} et~al.,}{{Gendreau}
  et~al.}{2016}]{NICER}
{Gendreau} K.~C.,  et~al., 2016, in {den Herder} J.-W.~A.,  {Takahashi} T.,
  {Bautz} M.,  eds,  Society of Photo-Optical Instrumentation Engineers (SPIE)
  Conference Series Vol. 9905, Space Telescopes and Instrumentation 2016:
  Ultraviolet to Gamma Ray. p. 99051H, \mn@doi{10.1117/12.2231304}

\bibitem[\protect\citeauthoryear{{Guillot} \& {Rutledge}}{{Guillot} \&
  {Rutledge}}{2014}]{Guillot:2014}
{Guillot} S.,  {Rutledge} R.~E.,  2014, \mn@doi [\apjl]
  {10.1088/2041-8205/796/1/L3}, \href
  {https://ui.adsabs.harvard.edu/abs/2014ApJ...796L...3G} {796, L3}

\bibitem[\protect\citeauthoryear{{G{\"u}ver}, {Wroblewski}, {Camarota}  \&
  {{\"O}zel}}{{G{\"u}ver} et~al.}{2010}]{Guver:2010b}
{G{\"u}ver} T.,  {Wroblewski} P.,  {Camarota} L.,   {{\"O}zel} F.,  2010,
  \mn@doi [\apj] {10.1088/0004-637X/719/2/1807}, \href
  {https://ui.adsabs.harvard.edu/abs/2010ApJ...719.1807G} {719, 1807}

\bibitem[\protect\citeauthoryear{Hebeler}{Hebeler}{2021}]{Hebeler:2020}
Hebeler K.,  2021, \mn@doi [Phys. Rept.] {10.1016/j.physrep.2020.08.009}, 890,
  1

\bibitem[\protect\citeauthoryear{{Heyl}}{{Heyl}}{2004}]{Heyl:2004}
{Heyl} J.~S.,  2004, \mn@doi [\apj] {10.1086/379966}, \href
  {https://ui.adsabs.harvard.edu/abs/2004ApJ...600..939H} {600, 939}

\bibitem[\protect\citeauthoryear{{Jahoda}, {Swank}, {Giles}, {Stark},
  {Strohmayer}, {Zhang}  \& {Morgan}}{{Jahoda} et~al.}{1996}]{Jahoda:1996}
{Jahoda} K.,  {Swank} J.~H.,  {Giles} A.~B.,  {Stark} M.~J.,  {Strohmayer} T.,
  {Zhang} W.,   {Morgan} E.~H.,  1996, in {Siegmund} O.~H.,  {Gummin} M.~A.,
  eds,  Society of Photo-Optical Instrumentation Engineers (SPIE) Conference
  Series Vol. 2808, EUV, X-Ray, and Gamma-Ray Instrumentation for Astronomy
  VII. pp 59--70, \mn@doi{10.1117/12.256034}

\bibitem[\protect\citeauthoryear{Kini et~al.,}{Kini
  et~al.}{2023a}]{zenodo_paper1}
Kini Y.,  et~al., 2023a, {Pulse Profile Modeling of Thermonuclear Burst
  Oscillations I: The Effect of Neglecting Variability},
  \mn@doi{10.5281/zenodo.7665653}, \url
  {https://doi.org/10.5281/zenodo.7665653}

\bibitem[\protect\citeauthoryear{Kini et~al.,}{Kini
  et~al.}{2023b}]{zenodo_paper2}
Kini Y.,  et~al., 2023b, {Pulse Profile Modelling of Thermonuclear Burst
  Oscillations II: Handling variability}, \mn@doi{10.5281/zenodo.8033527}, \url
  {https://doi.org/10.5281/zenodo.8033527}

\bibitem[\protect\citeauthoryear{{Kini} et~al.,}{{Kini}
  et~al.}{2023c}]{kini:2023}
{Kini} Y.,  et~al., 2023c, \mn@doi [\mnras] {10.1093/mnras/stad1030}, \href
  {https://ui.adsabs.harvard.edu/abs/2023MNRAS.522.3389K} {522, 3389}

\bibitem[\protect\citeauthoryear{Krauss et~al.,}{Krauss
  et~al.}{2005}]{Krauss:2005sj}
Krauss M.~I.,  et~al., 2005, \mn@doi [\apj] {10.1086/430595}, 627, 910

\bibitem[\protect\citeauthoryear{Lattimer}{Lattimer}{2012}]{Lattimer:2012}
Lattimer J.~M.,  2012, \mn@doi [Ann. Rev. Nucl. Part. Sci.]
  {10.1146/annurev-nucl-102711-095018}, 62, 485

\bibitem[\protect\citeauthoryear{{Lee}}{{Lee}}{2004}]{Lee:2004}
{Lee} U.,  2004, \mn@doi [\apj] {10.1086/380122}, \href
  {https://ui.adsabs.harvard.edu/abs/2004ApJ...600..914L} {600, 914}

\bibitem[\protect\citeauthoryear{Lo, Coleman~Miller, Bhattacharyya  \& Lamb}{Lo
  et~al.}{2013}]{Lo:2013ava}
Lo K.~H.,  Coleman~Miller M.,  Bhattacharyya S.,   Lamb F.~K.,  2013, \mn@doi
  [\apj] {10.1088/0004-637X/776/1/19}, 776, 19

\bibitem[\protect\citeauthoryear{{Mahmoodifar} \& {Strohmayer}}{{Mahmoodifar}
  \& {Strohmayer}}{2016}]{Mahmoodifar:2016}
{Mahmoodifar} S.,  {Strohmayer} T.,  2016, \mn@doi [\apj]
  {10.3847/0004-637X/818/1/93}, \href
  {https://ui.adsabs.harvard.edu/abs/2016ApJ...818...93M} {818, 93}

\bibitem[\protect\citeauthoryear{Meisel}{Meisel}{2018}]{Meisel:2018rsy}
Meisel Z.,  2018, \mn@doi [\apj] {10.3847/1538-4357/aac3d3}, 860, 147

\bibitem[\protect\citeauthoryear{{Miller} \& {Lamb}}{{Miller} \&
  {Lamb}}{1998}]{Miller:1998}
{Miller} M.~C.,  {Lamb} F.~K.,  1998, \mn@doi [\apjl] {10.1086/311335}, \href
  {https://ui.adsabs.harvard.edu/abs/1998ApJ...499L..37M} {499, L37}

\bibitem[\protect\citeauthoryear{{Miller} \& {Lamb}}{{Miller} \&
  {Lamb}}{2015}]{Miller:2015}
{Miller} M.~C.,  {Lamb} F.~K.,  2015, \mn@doi [\apj]
  {10.1088/0004-637X/808/1/31}, \href
  {https://ui.adsabs.harvard.edu/abs/2015ApJ...808...31M} {808, 31}

\bibitem[\protect\citeauthoryear{{Miller} et~al.,}{{Miller}
  et~al.}{2019}]{Miller:2019}
{Miller} M.~C.,  et~al., 2019, \mn@doi [\apjl] {10.3847/2041-8213/ab50c5},
  \href {https://ui.adsabs.harvard.edu/abs/2019ApJ...887L..24M} {887, L24}

\bibitem[\protect\citeauthoryear{{Miller} et~al.,}{{Miller}
  et~al.}{2021}]{Miller:2021}
{Miller} M.~C.,  et~al., 2021, \mn@doi [\apjl] {10.3847/2041-8213/ac089b},
  \href {https://ui.adsabs.harvard.edu/abs/2021ApJ...918L..28M} {918, L28}

\bibitem[\protect\citeauthoryear{{Morsink}, {Leahy}, {Cadeau}  \&
  {Braga}}{{Morsink} et~al.}{2007}]{Morsink:2007}
{Morsink} S.~M.,  {Leahy} D.~A.,  {Cadeau} C.,   {Braga} J.,  2007, \mn@doi
  [\apj] {10.1086/518648}, \href
  {https://ui.adsabs.harvard.edu/abs/2007ApJ...663.1244M} {663, 1244}

\bibitem[\protect\citeauthoryear{{Muno}, {Chakrabarty}, {Galloway}  \&
  {Psaltis}}{{Muno} et~al.}{2002a}]{muno02a}
{Muno} M.~P.,  {Chakrabarty} D.,  {Galloway} D.~K.,   {Psaltis} D.,  2002a,
  \mn@doi [\apj] {10.1086/343793}, \href
  {https://ui.adsabs.harvard.edu/abs/2002ApJ...580.1048M} {580, 1048}

\bibitem[\protect\citeauthoryear{{Muno}, {{\"O}zel}  \& {Chakrabarty}}{{Muno}
  et~al.}{2002b}]{Muno:2002}
{Muno} M.~P.,  {{\"O}zel} F.,   {Chakrabarty} D.,  2002b, \mn@doi [\apj]
  {10.1086/344152}, \href
  {https://ui.adsabs.harvard.edu/abs/2002ApJ...581..550M} {581, 550}

\bibitem[\protect\citeauthoryear{{N{\"a}ttil{\"a}} \&
  {Pihajoki}}{{N{\"a}ttil{\"a}} \& {Pihajoki}}{2018}]{Nattila:2018}
{N{\"a}ttil{\"a}} J.,  {Pihajoki} P.,  2018, \mn@doi [\aap]
  {10.1051/0004-6361/201630261}, \href
  {https://ui.adsabs.harvard.edu/abs/2018A&A...615A..50N} {615, A50}

\bibitem[\protect\citeauthoryear{{N{\"a}ttil{\"a}}, {Steiner}, {Kajava},
  {Suleimanov}  \& {Poutanen}}{{N{\"a}ttil{\"a}} et~al.}{2016}]{Nattila:2016}
{N{\"a}ttil{\"a}} J.,  {Steiner} A.~W.,  {Kajava} J.~J.~E.,  {Suleimanov}
  V.~F.,   {Poutanen} J.,  2016, \mn@doi [\aap] {10.1051/0004-6361/201527416},
  \href {https://ui.adsabs.harvard.edu/abs/2016A&A...591A..25N} {591, A25}

\bibitem[\protect\citeauthoryear{Oertel, Hempel, Kl\"ahn  \& Typel}{Oertel
  et~al.}{2017}]{Oertel:2016}
Oertel M.,  Hempel M.,  Kl\"ahn T.,   Typel S.,  2017, \mn@doi [Rev. Mod.
  Phys.] {10.1103/RevModPhys.89.015007}, 89, 015007

\bibitem[\protect\citeauthoryear{{{\"O}zel} \& {Psaltis}}{{{\"O}zel} \&
  {Psaltis}}{2009}]{Ozel:2009b}
{{\"O}zel} F.,  {Psaltis} D.,  2009, \mn@doi [\prd]
  {10.1103/PhysRevD.80.103003}, \href
  {https://ui.adsabs.harvard.edu/abs/2009PhRvD..80j3003O} {80, 103003}

\bibitem[\protect\citeauthoryear{{{\"O}zel}, {G{\"u}ver}  \&
  {Psaltis}}{{{\"O}zel} et~al.}{2009}]{Ozel:2009}
{{\"O}zel} F.,  {G{\"u}ver} T.,   {Psaltis} D.,  2009, \mn@doi [\apj]
  {10.1088/0004-637X/693/2/1775}, \href
  {https://ui.adsabs.harvard.edu/abs/2009ApJ...693.1775O} {693, 1775}

\bibitem[\protect\citeauthoryear{{{\"O}zel}, {Psaltis}, {G{\"u}ver}, {Baym},
  {Heinke}  \& {Guillot}}{{{\"O}zel} et~al.}{2016}]{Ozel:2016}
{{\"O}zel} F.,  {Psaltis} D.,  {G{\"u}ver} T.,  {Baym} G.,  {Heinke} C.,
  {Guillot} S.,  2016, \mn@doi [\apj] {10.3847/0004-637X/820/1/28}, \href
  {https://ui.adsabs.harvard.edu/abs/2016ApJ...820...28O} {820, 28}

\bibitem[\protect\citeauthoryear{{Page}}{{Page}}{1995}]{Page:1995}
{Page} D.,  1995, \mn@doi [\apj] {10.1086/175439}, \href
  {https://ui.adsabs.harvard.edu/abs/1995ApJ...442..273P} {442, 273}

\bibitem[\protect\citeauthoryear{{Pechenick}, {Ftaclas}  \&
  {Cohen}}{{Pechenick} et~al.}{1983}]{Pechenick:1983}
{Pechenick} K.~R.,  {Ftaclas} C.,   {Cohen} J.~M.,  1983, \mn@doi [\apj]
  {10.1086/161498}, \href
  {https://ui.adsabs.harvard.edu/abs/1983ApJ...274..846P} {274, 846}

\bibitem[\protect\citeauthoryear{{Piro} \& {Bildsten}}{{Piro} \&
  {Bildsten}}{2005}]{Piro:2005}
{Piro} A.~L.,  {Bildsten} L.,  2005, \mn@doi [\apj] {10.1086/430777}, \href
  {https://ui.adsabs.harvard.edu/abs/2005ApJ...629..438P} {629, 438}

\bibitem[\protect\citeauthoryear{{Poutanen} \& {Beloborodov}}{{Poutanen} \&
  {Beloborodov}}{2006}]{Poutanen:2006}
{Poutanen} J.,  {Beloborodov} A.~M.,  2006, \mn@doi [\mnras]
  {10.1111/j.1365-2966.2006.11088.x}, \href
  {https://ui.adsabs.harvard.edu/abs/2006MNRAS.373..836P} {373, 836}

\bibitem[\protect\citeauthoryear{Psaltis, \"Ozel  \& Chakrabarty}{Psaltis
  et~al.}{2014}]{Psaltis:2013fha}
Psaltis D.,  \"Ozel F.,   Chakrabarty D.,  2014, \mn@doi [\apj]
  {10.1088/0004-637X/787/2/136}, 787, 136

\bibitem[\protect\citeauthoryear{{Ray} et~al.,}{{Ray} et~al.}{2019}]{Ray:2019}
{Ray} P.~S.,  et~al., 2019, \mn@doi [arXiv e-prints]
  {10.48550/arXiv.1903.03035}, \href
  {https://ui.adsabs.harvard.edu/abs/2019arXiv190303035R} {p. arXiv:1903.03035}

\bibitem[\protect\citeauthoryear{{Riley}}{{Riley}}{2019}]{riley_thesis}
{Riley} T.~E.,  2019, PhD thesis, University of Amsterdam, \url
  {https://hdl.handle.net/11245.1/aa86fcf3-2437-4bc2-810e-cf9f30a98f7a}

\bibitem[\protect\citeauthoryear{{Riley} et~al.,}{{Riley}
  et~al.}{2019}]{Riley:2019}
{Riley} T.~E.,  et~al., 2019, \mn@doi [\apjl] {10.3847/2041-8213/ab481c}, \href
  {https://ui.adsabs.harvard.edu/abs/2019ApJ...887L..21R} {887, L21}

\bibitem[\protect\citeauthoryear{{Riley} et~al.,}{{Riley}
  et~al.}{2021}]{Riley:2021}
{Riley} T.~E.,  et~al., 2021, \mn@doi [\apjl] {10.3847/2041-8213/ac0a81}, \href
  {https://ui.adsabs.harvard.edu/abs/2021ApJ...918L..27R} {918, L27}

\bibitem[\protect\citeauthoryear{Riley et~al.,}{Riley et~al.}{2023}]{Riley2023}
Riley T.~E.,  et~al., 2023, \mn@doi [Journal of Open Source Software]
  {10.21105/joss.04977}, 8, 4977

\bibitem[\protect\citeauthoryear{{Salmi}, {N{\"a}ttil{\"a}}  \&
  {Poutanen}}{{Salmi} et~al.}{2018}]{Salmi:2018}
{Salmi} T.,  {N{\"a}ttil{\"a}} J.,   {Poutanen} J.,  2018, \mn@doi [\aap]
  {10.1051/0004-6361/201833348}, \href
  {https://ui.adsabs.harvard.edu/abs/2018A&A...618A.161S} {618, A161}

\bibitem[\protect\citeauthoryear{{Salmi} et~al.,}{{Salmi}
  et~al.}{2022}]{Salmi:2022}
{Salmi} T.,  et~al., 2022, \mn@doi [\apj] {10.3847/1538-4357/ac983d}, \href
  {https://ui.adsabs.harvard.edu/abs/2022ApJ...941..150S} {941, 150}

\bibitem[\protect\citeauthoryear{{Spitkovsky}, {Levin}  \&
  {Ushomirsky}}{{Spitkovsky} et~al.}{2002}]{Spitkovsky:2002}
{Spitkovsky} A.,  {Levin} Y.,   {Ushomirsky} G.,  2002, \mn@doi [\apj]
  {10.1086/338040}, \href
  {https://ui.adsabs.harvard.edu/abs/2002ApJ...566.1018S} {566, 1018}

\bibitem[\protect\citeauthoryear{{Steiner}, {Lattimer}  \& {Brown}}{{Steiner}
  et~al.}{2013}]{Steiner:2013}
{Steiner} A.~W.,  {Lattimer} J.~M.,   {Brown} E.~F.,  2013, \mn@doi [\apjl]
  {10.1088/2041-8205/765/1/L5}, \href
  {https://ui.adsabs.harvard.edu/abs/2013ApJ...765L...5S} {765, L5}

\bibitem[\protect\citeauthoryear{{Steiner}, {Heinke}, {Bogdanov}, {Li}, {Ho},
  {Bahramian}  \& {Han}}{{Steiner} et~al.}{2018}]{Steiner:2018}
{Steiner} A.~W.,  {Heinke} C.~O.,  {Bogdanov} S.,  {Li} C.~K.,  {Ho} W.~C.~G.,
  {Bahramian} A.,   {Han} S.,  2018, \mn@doi [\mnras] {10.1093/mnras/sty215},
  \href {https://ui.adsabs.harvard.edu/abs/2018MNRAS.476..421S} {476, 421}

\bibitem[\protect\citeauthoryear{{Stevens}, {Fiege}, {Leahy}  \&
  {Morsink}}{{Stevens} et~al.}{2016}]{Stevens:2016}
{Stevens} A.~L.,  {Fiege} J.~D.,  {Leahy} D.~A.,   {Morsink} S.~M.,  2016,
  \mn@doi [\apj] {10.3847/1538-4357/833/2/244}, \href
  {https://ui.adsabs.harvard.edu/abs/2016ApJ...833..244S} {833, 244}

\bibitem[\protect\citeauthoryear{{Strohmayer}, {Zhang}, {Swank}, {Smale},
  {Titarchuk}, {Day}  \& {Lee}}{{Strohmayer} et~al.}{1996}]{Strohmayer:1996}
{Strohmayer} T.~E.,  {Zhang} W.,  {Swank} J.~H.,  {Smale} A.,  {Titarchuk} L.,
  {Day} C.,   {Lee} U.,  1996, \mn@doi [\apjl] {10.1086/310261}, \href
  {https://ui.adsabs.harvard.edu/abs/1996ApJ...469L...9S} {469, L9}

\bibitem[\protect\citeauthoryear{{Strohmayer}, {Zhang}  \&
  {Swank}}{{Strohmayer} et~al.}{1997}]{Strohmayer:1997}
{Strohmayer} T.~E.,  {Zhang} W.,   {Swank} J.~H.,  1997, \mn@doi [\apjl]
  {10.1086/310880}, \href
  {https://ui.adsabs.harvard.edu/abs/1997ApJ...487L..77S} {487, L77}

\bibitem[\protect\citeauthoryear{Strohmayer, Markwardt, Swank  \& in~'t
  Zand}{Strohmayer et~al.}{2003}]{Strohmayer:2003}
Strohmayer T.~E.,  Markwardt C.~B.,  Swank J.~H.,   in~'t Zand J.,  2003,
  \mn@doi [\apjl] {10.1086/379158}, 596, L67

\bibitem[\protect\citeauthoryear{{Suleimanov}, {Poutanen}, {Revnivtsev}  \&
  {Werner}}{{Suleimanov} et~al.}{2011}]{Suleimanov:2011}
{Suleimanov} V.,  {Poutanen} J.,  {Revnivtsev} M.,   {Werner} K.,  2011,
  \mn@doi [\apj] {10.1088/0004-637X/742/2/122}, \href
  {https://ui.adsabs.harvard.edu/abs/2011ApJ...742..122S} {742, 122}

\bibitem[\protect\citeauthoryear{{Suleimanov}, {Poutanen}  \&
  {Werner}}{{Suleimanov} et~al.}{2012}]{Valery2012}
{Suleimanov} V.,  {Poutanen} J.,   {Werner} K.,  2012, \mn@doi [\aap]
  {10.1051/0004-6361/201219480}, \href
  {https://ui.adsabs.harvard.edu/abs/2012A&A...545A.120S} {545, A120}

\bibitem[\protect\citeauthoryear{Tolos \& Fabbietti}{Tolos \&
  Fabbietti}{2020}]{Tolos:2020}
Tolos L.,  Fabbietti L.,  2020, \mn@doi [Prog. Part. Nucl. Phys.]
  {10.1016/j.ppnp.2020.103770}, 112, 103770

\bibitem[\protect\citeauthoryear{{Vaughan}, {Edelson}, {Warwick}  \&
  {Uttley}}{{Vaughan} et~al.}{2003}]{Vaughan}
{Vaughan} S.,  {Edelson} R.,  {Warwick} R.~S.,   {Uttley} P.,  2003, \mn@doi
  [\mnras] {10.1046/j.1365-2966.2003.07042.x}, \href
  {https://ui.adsabs.harvard.edu/abs/2003MNRAS.345.1271V} {345, 1271}

\bibitem[\protect\citeauthoryear{{Vinciguerra}, {Salmi}, {Watts}, {Choudhury},
  {Kini}  \& {Riley}}{{Vinciguerra} et~al.}{2023}]{vinciguerra:2023_sim}
{Vinciguerra} S.,  {Salmi} T.,  {Watts} A.~L.,  {Choudhury} D.,  {Kini} Y.,
  {Riley} T.~E.,  2023, ApJ, submitted

\bibitem[\protect\citeauthoryear{{Wang}, {Steeghs}, {Casares}, {Charles},
  {Mu{\~n}oz-Darias}, {Marsh}, {Hynes}  \& {O'Brien}}{{Wang}
  et~al.}{2017}]{Wang:2017}
{Wang} L.,  {Steeghs} D.,  {Casares} J.,  {Charles} P.~A.,  {Mu{\~n}oz-Darias}
  T.,  {Marsh} T.~R.,  {Hynes} R.~I.,   {O'Brien} K.,  2017, \mn@doi [\mnras]
  {10.1093/mnras/stw3312}, \href
  {https://ui.adsabs.harvard.edu/abs/2017MNRAS.466.2261W} {466, 2261}

\bibitem[\protect\citeauthoryear{{Watts}, {Patruno}  \& {van der Klis}}{{Watts}
  et~al.}{2008}]{Watts:2008}
{Watts} A.~L.,  {Patruno} A.,   {van der Klis} M.,  2008, \mn@doi [\apjl]
  {10.1086/594365}, \href
  {https://ui.adsabs.harvard.edu/abs/2008ApJ...688L..37W} {688, L37}

\bibitem[\protect\citeauthoryear{{Watts} et~al.,}{{Watts}
  et~al.}{2016}]{Watts:2016}
{Watts} A.~L.,  et~al., 2016, \mn@doi [Reviews of Modern Physics]
  {10.1103/RevModPhys.88.021001}, \href
  {https://ui.adsabs.harvard.edu/abs/2016RvMP...88b1001W} {88, 021001}

\bibitem[\protect\citeauthoryear{{Watts} et~al.,}{{Watts}
  et~al.}{2019}]{Watts:2019_extp}
{Watts} A.~L.,  et~al., 2019, \mn@doi [Science China Physics, Mechanics, and
  Astronomy] {10.1007/s11433-017-9188-4}, \href
  {https://ui.adsabs.harvard.edu/abs/2019SCPMA..6229503W} {62, 29503}

\bibitem[\protect\citeauthoryear{{Weinberg}, {Miller}  \& {Lamb}}{{Weinberg}
  et~al.}{2001}]{Weinberg:2001}
{Weinberg} N.,  {Miller} M.~C.,   {Lamb} D.~Q.,  2001, \mn@doi [\apj]
  {10.1086/318279}, \href
  {https://ui.adsabs.harvard.edu/abs/2001ApJ...546.1098W} {546, 1098}

\bibitem[\protect\citeauthoryear{{Wilms}, {Allen}  \& {McCray}}{{Wilms}
  et~al.}{2000}]{Wilms:2000}
{Wilms} J.,  {Allen} A.,   {McCray} R.,  2000, \mn@doi [\apj] {10.1086/317016},
  \href {https://ui.adsabs.harvard.edu/abs/2000ApJ...542..914W} {542, 914}

\bibitem[\protect\citeauthoryear{Yang \& Piekarewicz}{Yang \&
  Piekarewicz}{2020}]{Yang:2019}
Yang J.,  Piekarewicz J.,  2020, \mn@doi [Ann. Rev. Nucl. Part. Sci.]
  {10.1146/annurev-nucl-101918-023608}, 70, 21

\bibitem[\protect\citeauthoryear{{Zhang} et~al.,}{{Zhang}
  et~al.}{2019}]{Zhang:2019}
{Zhang} S.,  et~al., 2019, \mn@doi [Science China Physics, Mechanics, and
  Astronomy] {10.1007/s11433-018-9309-2}, \href
  {https://ui.adsabs.harvard.edu/abs/2019SCPMA..6229502Z} {62, 29502}

\makeatother
\end{thebibliography}



\begin{appendix}


\section{Appendix }\label{sec:appendix}

\subsection{Total likelihood derivation}\label{sec:like_deriv}

Let denote by $\mathcal{D}= \{d_i\}_{i=1}^{K}$ the complete dataset, where $d_i$ corresponds to the data for the $i^{\mathrm{th}}$ segment and $K$ represents the total number of segments ($K=8$). Let ($\theta, \phi_i$) represent the model vector for model $m_i$ used to model data $d_i$, where $\theta$ is the shared parameter vector for all models $\mathcal{M}= \{m_i\}_{i=1}^{K}$and $\phi_i$ is the specific parameter vector for model $m_i$. The total likelihood can be defined as:

\begin{equation}\label{eq}
\begin{split}
   \mathcal{L}_{\mathrm{total}}  & \stackrel{\mathrm{def}}{=} p(\{d_i\}_{i=1}^{K}|\theta,\phi_1,\phi_2, ..., \phi_K)  \\
                                                           & =p(d_1, d_2, ...,d_K|\theta,\phi_1,\phi_2, ..., \phi_K)  \\
                                                           & =\frac{p(d_1, d_2, ...,d_K,\theta,\phi_1,\phi_2, ..., \phi_K)}{p(\theta,\phi_1,\phi_2, ..., \phi_K)}.
\end{split}
\end{equation}

We can further simplify the above expression by using the conditional independence property of the segments given the parameters. To simplify the writing, let us denote by $d_0$ =$\{\theta,\phi_1,\phi_2,..., \phi_K \}$. From the chain rule, we have:

\begin{equation}
\begin{split}
    p(d_0,d_1, d_2, ...,d_K) &= p(d_0) \prod_{i=1}^{K}p(d_i|d_0, ...,d_{i-1}).\\
\end{split}
\end{equation}

But, 

\begin{equation}
\begin{split}
   p(d_i|d_0, ...,d_{i-1}) &= p(d_i| \theta,\phi_1,\phi_2,..., \phi_K, ...,d_{i-1}) \\
                        &= p(d_i| \theta,\phi_i). \end{split}
\end{equation}

Hence, 

\begin{equation}\label{eq_f}
\begin{split}
    \mathcal{L}_{\mathrm{total}} &= \prod_{i=1}^{K} p(d_{i}|\theta,\phi_i)  \\
                        & = \prod_{i=1}^{K} \mathcal{L}_{i}.
\end{split}
\end{equation}

\subsection{Additional figures}\label{sec:add_figures}
\begin{figure}
    \centering
    \includegraphics[width=\columnwidth]{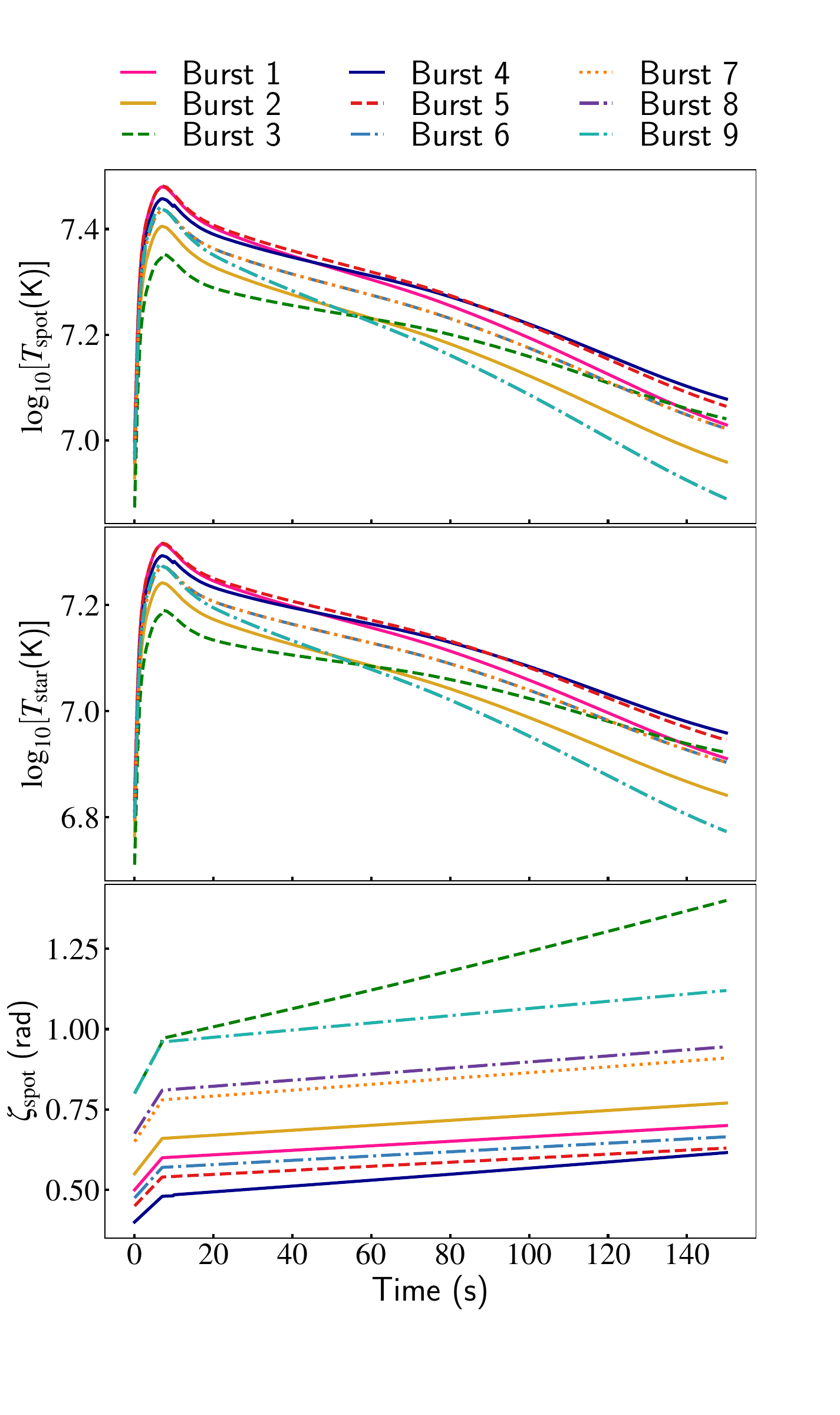}
      \caption{Parameters evolution used to generate the data of the second data subset.}
    \label{fig:profiles}
\end{figure}

\begin{figure}
    \centering
    \includegraphics[width=\columnwidth]{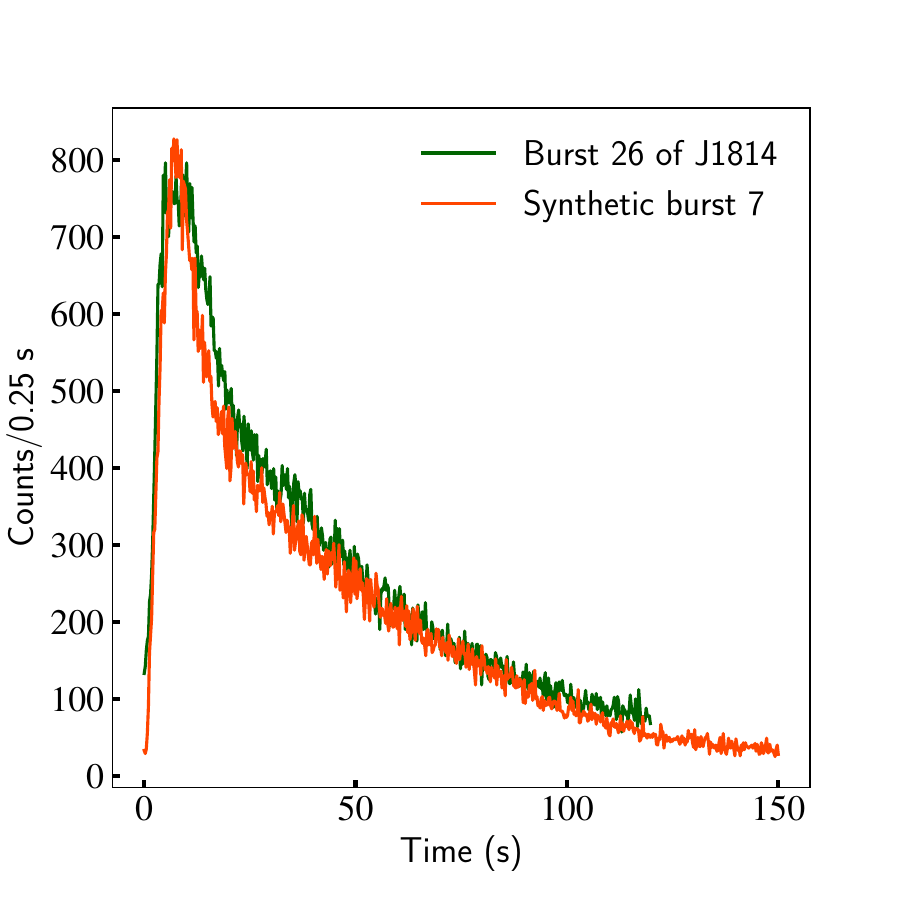}
      \caption{ Typical burst of J1814 (Burst 26, ObsID:80418-01-05-03) compared to Burst 7 of the second data subset.}
    \label{fig:example_1814_vs_synthetic}
\end{figure}

Figures \ref{fig:profiles} and \ref{fig:example_1814_vs_synthetic} present respectively the volutions used for generating the second data subset and the comparison between the light curve of Burst 7 from that data subset to a typical burst of J1814 (Burst 26, ObsID:80418-01-05-03).

\FloatBarrier

\subsection{Priors}\label{sec:priors}

Table \ref{tab:Priors} shows a comparison of the prior distributions for each parameter between the BF and WM methods for burst shown in Figure \ref{fig:profile_example}. It is apparent that the WM method exhibits a substantial reduction in the prior space, resulting in significant improvements in computing time.

\subsection{Hardware specifications}\label{sec:specs}
Some of the  key hardware and operating system specifications  are presented in Table \ref{tab:specs}. The links in the caption of the table lead to a more detailed description of each hardware component.

\onecolumn

\clearpage

\begin{deluxetable}{lll}
\tabletypesize{\small}

\tablewidth{0pt}

\tablecaption{Comparison of the priors distributions of each parameter for both the BF and the WM method for the burst shown in Figure \ref{fig:profile_example}\label{tab:Priors}}

\tablehead{\colhead{Parameters}\hspace{3.75cm}  & \colhead{\hspace{1.5cm}BF}\hspace{3.75cm} & \colhead{\hspace{.5cm}wM} \\}
\tablecolumns{3}
\startdata
\sidehead{\textbf{Shared across segments}}

 $M$ ($M_{\odot}$)  & $M\sim\mathcal{U}(1.0,3.0)$  & $M\sim\mathcal{U}(1.0,3.0)$  \\
$R_{\rm eq}$ (km)  &  $R_{\rm eq}\sim\mathcal{U}(3r_g(1.0),16.0)$\tablenotemark{a} & $R_{\rm eq}\sim\mathcal{U}(3r_{\mathrm{g}}(1.0),16.0)$  \\
$D$ (kpc)\tablenotemark{b}           &$D\sim\mathcal{N}(5.30,0.79)$  & $D\sim\mathcal{N}(5.30,0.79)$  \\
$\cos(i)$  & $\cos(i)\sim\mathcal{U}(0.0,1.0)$ & $\cos(i)\sim\mathcal{U}(0.0,1.0)$  \\
$\phi_\mathrm{spot}$ (cycles)   & $\phi_\mathrm{spot}\sim\mathcal{U}(-0.25,0.75)$ &$\phi_\mathrm{spot}\sim\mathcal{U}(-0.25,0.75)$  \\
$\Theta_\mathrm{spot}$ (rad)  &$\Theta_\mathrm{spot} \sim\mathcal{U}(0.0,\pi/2)$ & $\Theta_\mathrm{spot} \sim\mathcal{U}(0.0,\pi/2)$  \\
$T_\mathrm{els}\equiv \log_{10}[T_\mathrm{star}(\mathrm{K})/1\mathrm{K}]$   & $T_\mathrm{els} \sim\mathcal{U}(6.7,7.6)$ & $T_\mathrm{els} \sim\mathcal{U}(6.7,7.6)$ \\
$N_{H}$   ($10^{20}\mathrm{cm}^{-2}$)   & $N_{H} \sim \mathcal{U}(0.0,10.0)$  & $N_{H} \sim \mathcal{U}(0.0,10.0)$  \\

\sidehead{\textbf{Hot spot} ($T_\mathrm{spotX}\equiv \log_{10}[T_\mathrm{spotX}(\mathrm{K})/1\mathrm{K}]$)}

$T_\mathrm{spot1}$ & $T_\mathrm{spot1} \sim\mathcal{U}(6.7,7.6)$  &$T_\mathrm{spot1} \sim\mathcal{U}(7.13,7.26)$  \\
$T_\mathrm{spot2}$ & $T_\mathrm{spot2} \sim\mathcal{U}(6.7,7.6)$  &$T_\mathrm{spot2} \sim\mathcal{U}(7.26,7.30)$  \\
$T_\mathrm{spot3}$ & $T_\mathrm{spot3} \sim\mathcal{U}(6.7,7.6)$  &$T_\mathrm{spot3} \sim\mathcal{U}(7.28,7.32)$  \\
$T_\mathrm{spot4}$ & $T_\mathrm{spot4} \sim\mathcal{U}(6.7,7.6)$  &$T_\mathrm{spot4} \sim\mathcal{U}(7.22,7.28)$  \\
$T_\mathrm{spot5}$ & $T_\mathrm{spot5} \sim\mathcal{U}(6.7,7.6)$  &$T_\mathrm{spot5} \sim\mathcal{U}(7.18,7.22)$  \\
$T_\mathrm{spot6}$ & $T_\mathrm{spot6} \sim\mathcal{U}(6.7,7.6)$  &$T_\mathrm{spot6} \sim\mathcal{U}(7.14,7.18)$  \\
$T_\mathrm{spot7}$ & $T_\mathrm{spot7} \sim\mathcal{U}(6.7,7.6)$  &$T_\mathrm{spot7} \sim\mathcal{U}(7.09,7.14)$  \\
$T_\mathrm{spot8}$ & $T_\mathrm{spot8} \sim\mathcal{U}(6.7,7.6)$  &$T_\mathrm{spot8} \sim\mathcal{U}(7.04,7.09)$  \\

\sidehead{\textbf{Angular radius} ( in rad)} 

$\zeta_\mathrm{spot1}$ &  $\zeta_\mathrm{spot1}\sim\mathcal{U}(0.0,\pi/2)$ & $\zeta_\mathrm{spot1}\sim\mathcal{U}(0.71, 0.75)$\\
$\zeta_\mathrm{spot2}$ &  $\zeta_\mathrm{spot2}\sim\mathcal{U}(0.0,\pi/2)$ & $\zeta_\mathrm{spot2}\sim\mathcal{U}(0.75,0.76)$\\
$\zeta_\mathrm{spot3}$ &  $\zeta_\mathrm{spot3}\sim\mathcal{U}(0.0,\pi/2)$ & $\zeta_\mathrm{spot3}\sim\mathcal{U}(0.76,0.81)$ \\
$\zeta_\mathrm{spot4}$ &  $\zeta_\mathrm{spot4}\sim\mathcal{U}(0.0,\pi/2)$ & $\zeta_\mathrm{spot4}\sim\mathcal{U}(0.81,0.83)$\\
$\zeta_\mathrm{spot5}$ &  $\zeta_\mathrm{spot5}\sim\mathcal{U}(0.0,\pi/2)$ & $\zeta_\mathrm{spot5}\sim\mathcal{U}(0.83,0.86)$ \\
$\zeta_\mathrm{spot6}$ &  $\zeta_\mathrm{spot6}\sim\mathcal{U}(0.0,\pi/2)$ & $\zeta_\mathrm{spot6}\sim\mathcal{U}(0.86,0.89)$\\
$\zeta_\mathrm{spot7}$ &  $\zeta_\mathrm{spot7}\sim\mathcal{U}(0.0,\pi/2)$ & $\zeta_\mathrm{spot7}\sim\mathcal{U}(0.89,0.91)$\\
$\zeta_\mathrm{spot8}$ &  $\zeta_\mathrm{spot8}\sim\mathcal{U}(0.0,\pi/2)$ & $\zeta_\mathrm{spot8}\sim\mathcal{U}(0.91,0.94)$\\

\enddata
\tablenotetext{a}{$r_{\mathrm{g}}(1.0)$: Solar Schwarzschild gravitational radius}
\tablenotetext{b}{We set the distance prior as $D \sim \mathcal{N}(d, 0.15d)$, where $d$ is the distance used to generate the synthetic data, as in \citetalias{kini:2023}.}

\end{deluxetable}

\clearpage

\begin{deluxetable}{lll}
\tabletypesize{\small}

\tablewidth{0pt}

\tablecaption{Hardware  and operating system specifications\tablenotemark{*}}\label{tab:specs}

\tablehead{\colhead{}\hspace{3.75cm}  & \colhead{\hspace{1.5cm}HELIOS cluster}\hspace{3.75cm} & \colhead{\hspace{.5cm} Lisa cluster} \\}
\tablecolumns{3}
\startdata

Partition               & Neutron-star                                 & Gold\_6130  \\
Operating system        & CentOS 7                                     & Debian 10 \\
Model name              & AMD EPYC 7452 32-Core Processor              &Intel$^{\circledR}$ Xeon$^{\circledR}$ Gold 6130 Processor  \\
Number of CPU           & 2                                            & 1  \\
Number of cores per CPU & 32                                           & 16  \\
Total threads           & 128\tablenotemark{**}                         & 32\tablenotemark{**} \\
Base frequency          & 2.35 GHz                                     &2.10 GHz \\

\enddata
\tablenotetext{*}{We only provided the general specifications. For the full specifications of the hardware, see \url{https://www.amd.com/en/product/8801} and \url{https://ark.intel.com/content/www/us/en/ark/products/120492/intel-xeon-gold-6130-processor-22m-cache-2-10-ghz.html}  for respectively the HELIOS and Lisa cluster.}
\tablenotetext{**}{For a given run, the maximum number of threads that could be requested  was 126 and 16 respectively for the HELIOS and Lisa cluster.}
\end{deluxetable}
\end{appendix}

\bsp	
\label{lastpage}
\end{document}